\documentclass[preprintnumbers,amsmath,amssymb,aps]{revtex4}
\usepackage{graphicx}
\usepackage{latexsym}
\usepackage[dvips]{color}

\preprint{COLO-HEP-533}
\preprint{HU-EP-07/59}
\preprint{LPT-ORSAY 07-126}
\preprint{SFB/CPP-07-82}

\def\Tr{{\rm Tr}}

\begin{document}
\title{Diquark effects in light baryon correlators from lattice QCD}
\author{Thomas DeGrand}
\affiliation{
Department of Physics, University of Colorado,
        Boulder, CO 80309 USA}
\author{Zhaofeng Liu}
\affiliation{
Laboratoire de Physique Th\'eorique
(B\^at. 210), UMR8627, Universit\'e de
Paris XI et CNRS, Centre d'Orsay, 91405 Orsay-Cedex, France
}
\author{Stefan Schaefer}
\affiliation{
Institut f\"ur Physik, Humboldt-Universit\"at zu Berlin, Newtonstr. 15,
12489 Berlin, Germany
}

\begin{abstract}
We study the role of diquarks in light baryons through 
point to point baryon correlators. We contrast results from quenched simulations with ones
with
two flavors of dynamical overlap fermions. The scalar,
pseudoscalar and axial vector diquarks are combined with light quarks
to form color singlets. The quenched simulation shows large zero mode effects
in correlators containing the scalar and pseudoscalar diquark.
The two scalar diquarks created by $\gamma_5$ and $\gamma_0\gamma_5$
lead to different behavior in baryon correlators,
showing that the interaction of diquarks with the third light quark
matters: we do not see an isolated diquark.
In our quark mass range, the scalar diquark created by $\gamma_5$ seems 
to play a greater role than
the others.
\end{abstract}
\maketitle


\section{Introduction}\label{Intr}
In the conventional quark model, a baryon is a bound state of three more-or-less
equivalent valence quarks. However, there are variants of the quark model in which interactions 
give rise to a quark-diquark structure for the baryon, where the binding of two
of the quarks in the baryon is different from the bound state properties of the third quark.
For a review of diquark phenomenology,  see 
\cite{Anselmino:1992vg}.

One would like to understand the properties of diquark correlations  (if any)
directly from QCD. Since this is a problem of the long distance behavior 
of the strong interactions,
perturbation theory is not applicable. Lattice QCD is a potential source
of information about diquark structures, and the subject has been visited 
many times in lattice simulations 
\cite{Babich:2007ah,Babich:2006eu,Alexandrou:2006cq,Alexandrou:2005zn,Orginos:2005vr,Fodor:2005qx,Hess:1998sd}.

Diquarks are not color singlets. Many techniques have been used in lattice simulations
to deal with this fact.
In Refs.~\cite{Babich:2007ah,Hess:1998sd}, gauge dependent diquark
correlators were calculated in a fixed gauge to study the masses of
diquarks. 
In Refs.~\cite{Alexandrou:2006cq,Alexandrou:2005zn,Orginos:2005vr} gauge invariance was preserved by
investigating the properties of states containing a static quark coupled gauge-invariantly to
two light quarks. Their quantum numbers combine to give the quantum numbers of the diquark.

Whatever a diquark is, the environment it feels in
a baryon with a heavy spectator quark is presumably different from that in a baryon
with a third light quark.
In this work, we want to study diquark correlations in
light baryons.  The interaction between a diquark and a
light quark  probably depends on the spin of the diquark and this interaction
can affect the formation of the diquark in the light baryon.
Here we combine a light quark with
a diquark with a specific quantum number in the color anti-triplet
representation to
get a color singlet. 

This work addresses two questions: First, do baryon correlators involving light quarks
show qualitative features associated with diquarks? We attack this question using both
quenched simulations and data from simulations with two flavors of light dynamical fermions.
As the reader will see, we had to deal with
a second, different question along the way: are the properties of baryon correlators
affected by the presence of dynamical fermions? The answer to that question is Yes: with
the simulation parameters we used, the
quenched approximation is afflicted by artifacts which make it unsuitable
to look for diquark effects.

We look for diquark effects through the point to point
correlator for baryonic currents, using the methodology first
described by Shuryak \cite{Shuryak:1993kg}: we compare
the correlator to its free-field value. If the ratio of the full correlator
to its free field value exceeds unity, we say that the channel is attractive;
if it falls below unity we say that the channel is repulsive.
An unphysical situation which is encountered in quenched simulations is
that the ratio becomes negative. This indicates a violation of unitarity.

We are interested in diquarks in the color
anti-triplet representation since
diquarks in the color sextet representation have much larger color
electrostatic energy and thus are not favored by phenomenology~\cite{Jaffe:2004ph}.
The scalar diquark 
is often called a ``good" diquark since the two quarks
are thought to be attractive to each other. The $1^+$ diquark is called a
``bad" diquark because all models suggest that it is heavier than
the scalar diquark. Besides these two diquarks, we also study the
pseudoscalar diquark. In a quenched
simulation, the point to point scalar meson correlator
was found to be negative~\cite{DeGrand:2001tm}. This behavior, 
and similar behavior for the scalar correlator in other situations, is known to be a quenching artifact. 
Since quark-quark interactions resemble quark-antiquark interactions,
we are curious to see if we
will see similar quenching artifacts in the pseudoscalar diquark channel.

Excess zero modes of Dirac operators at small quark masses
are lattice artifacts in quenched lattice simulations.
In a preliminary version of this work, reported
in \cite{Liu:2006zi}, we found that behavior of baryon correlators was strongly affected by zero modes.
In the sector of our data which had no zero modes, the difference among different
 baryon correlators was much smaller.
In that study, we also had data from simulations with dynamical fermions, but
they were performed in sectors of fixed topology and could not be applied directly to
phenomenology.  Here we will give results from 
dynamical simulations in which topological charge is allowed to fluctuate.

The paper is organized as follows: In Section \ref{Meth}, we present our
set up and simulation details.
Numerical results and discussions are given in Section \ref{NumRes}. 
Then we conclude in Section \ref{Summ}.

\section{Methodology}\label{Meth}

\subsection{Lattice action and simulation details}

Our simulations are performed using overlap fermions.
We have two data sets: one is quenched and the other is from a simulation with
two degenerate flavors of light fermions. The simulations have many features in common, allowing for
a reasonably straightforward comparison of results.

In both simulations the ``kernel'' (nonchiral  Dirac operator $d$ inserted into the
overlap operator $D= r_0( 1 + (d-r_0)/\sqrt{|d-r_0|^2})$) contains nearest and next-nearest terms and
a clover term set to its tree level value. The actual couplings are slightly different;
the action used in the quenched simulations is described in Ref.~\cite{DeGrand:2000tf}
while the dynamical action is given in Ref.~\cite{DeGrand:2004nq}.

Both simulations use gauge connections which were blocked over a hypercube.
The quenched simulations employ the implementation
of Hasenfratz and Knechtli \cite{Hasenfratz:2001hp}
while the dynamical simulations used the differentiable hypercubic link of
Hasenfratz, Hoffmann, and Schaefer \cite{Hasenfratz:2007rf}.
The choice of blocking parameters, $\alpha_1=0.75$, $\alpha_2=0.6$, $\alpha_3=0.3$,
in the conventions of Ref. \cite{Hasenfratz:2001hp}, 
are identical. These two blockings produce identical results in perturbation theory.

Our quenched data set has 40 configurations with lattice size $16^4$
and were generated using the Wilson gauge action at a
gauge coupling $\beta=6.1$. The bare quark masses $am_q$ are 0.015, 0.025
and 0.05. The lattice spacing is $a=0.08$fm
determined from the Sommer parameter $r_0$, assuming a nominal physical value of 0.5 fm.
Thus the lattices have a physical size of about 1.3 fm, and the quark masses
are about 37, 60, and 120 MeV.

The quenched simulations exhibit a typical quenched topological susceptibility.
From the zero-mode spectrum, they have $\langle Q_T \rangle=-0.15(30)$ and 
$\langle Q_T^2 \rangle=3.55(61)$. This corresponds to a scaled topological susceptibility of
$r_0^4 \chi_T=0.086(15)$ or $\chi_T = (213(9)$ MeV)${}^4$, not too different from recent high-statistics
results \cite{Durr:2006ky}.

 In the two flavor dynamical simulation the lattice size is $12^4$. 
We used a tadpole-improved L\"uscher-Weisz action with a coupling
$\beta=7.3$ and performed simulations  at three quark masses: $am_q=0.03$, 0.05 and 0.10.
The Sommer parameters are found to be
$r_0/a=3.70(5)$ at quark mass 0.03, 3.49(4) at 0.05 and 3.39(3) at 0.10.
Thus the lattice spacing is about 0.14 fm,  the physical size of these lattices is $L \sim 1.7$ fm, and the quark masses are 43, 69, and 133 MeV.
The pion masses in lattice units are 
0.32, 0.43 and 0.59
respectively.
Details of the simulations are found in Ref. \cite{TDSSappear}.
Our measurements of the correlators are done on 60 to 80 configurations from these data sets.
The topological susceptibility shows a dramatic variation with the quark mass, roughly
consistent with the expected linear dependence $\chi_T = m_q \Sigma/2$.

A strength of our simulation is the use of a fermion with continuum-like chiral symmetry
at finite lattice spacing. It will turn out that the correlation functions we measure have a fair amount
of sensitivity to the underlying topology of the gauge configurations. Simply counting the number of 
chiral zero modes gives us a robust measure of the topology.

Another strength is the use of full QCD with identical valence and sea quarks, rather than
using simulations with different formulations of valence and sea quarks, or partially 
quenched simulations (i.e., with different valence and sea quark masses). This way we avoid
a host of artifacts associated with the peculiar chiral properties of such formulations.
The particular disease we want to avoid is associated with scalar correlators. This is best known
in the situation of scalar mesons, where there is a long history of observations
going back to the quenched studies of Ref. \cite{Bardeen:2001jm}.
Basically, the scalar meson correlator includes a pi-eta-prime
 intermediate state. In quenched or partially quenched simulations, the flavor singlet pseudoscalar
channel contains an unphysical double pole at finite lattice spacing.
This intermediate state can dominate the correlation function; removing it requires all the machinery
of partially quenched chiral perturbation theory (see Ref. \cite{Prelovsek:2005rf},
plus the recent summaries of Refs. \cite{Durr:2007ef,Aubin:2007wr}).

In addition,  zero modes of the Dirac operator (in sectors of nonzero topology)
make a negative contribution to the scalar correlator \cite{Edwards:1998wx}.
 This has been seen in point-to-point
meson correlators in small volume \cite{DeGrand:2001tm}.

The current algebra description for scalar mesons is of course lacking for 
pseudoscalar combinations
of two quarks, but the disease is still present. We will see negative correlation functions in our
quenched data. They are expected to be absent in our full
QCD simulations, and they will be.

An obvious weakness of our study is that the simulation volume is small.
We do not trust our measurements of baryon masses; while we can use a combination of
quark propagators which are
periodic and antiperiodic in time to go to long temporal separations in the correlation function,
 the three-dimensional volumes probably squeeze the baryon wave functions, affecting their
masses. In addition, as the quark mass falls to zero at fixed volume $V=L^4$, at some point $m_\pi L$
becomes order unity. At this point, the finite volume induces an effective restoration of chiral
symmetry; the condensate is pushed to zero. This so-called epsilon-regime is interesting in its
own right, but cannot be used to make direct connections to the properties of hadrons in infinite volume.
Clearly, any future simulation should be done with larger volume.

\subsection{Observables}
The baryon currents and
correlation functions we considered are collected in Table~\ref{currenttab}.
Here $C$ is the charge-conjugation operator.
\begin{table}
\caption{Currents and correlation functions}
\begin{center}
\begin{tabular}{ccll}
\hline\hline
$J^P$ (diquark) & Color & Current & Correlator $R(x)$ \\
\hline
$0^+$ & $\bar{3}$ & $J^5=\epsilon_{abc}[u^aC\gamma_5d^b]u^c$ &
$\frac{1}{4}\Tr[\langle\Omega|TJ^5(x)\bar{J}^5(0)|\Omega\rangle
x_\nu\gamma_\nu]$ \\
$0^+$ & $\bar{3}$ & $J^{05}=\epsilon_{abc}[u^aC\gamma_0\gamma_5d^b]u^c$
&$\frac{1}{4}\Tr[\langle\Omega|TJ^{05}(x)\bar{J}^{05}(0)|\Omega\rangle
x_\nu\gamma_\nu]$ \\
$0^-$ & $\bar{3}$ & $J^I=\epsilon_{abc}[u^aCd^b]u^c$ &
$\frac{1}{4}\Tr[\langle\Omega|TJ^I(x)\bar{J}^I(0)|\Omega\rangle
x_\nu\gamma_\nu]$ \\
$1^+$ & $\bar{3}$ & $J^3=\epsilon_{abc}[u^aC\gamma_3d^b]u^c$ &
$\frac{1}{4}\Tr[\langle\Omega|TJ^3(x)\bar{J}^3(0)|\Omega\rangle
x_\nu\gamma_\nu]$ \\
\hline\hline
\end{tabular}
\label{currenttab}
\end{center}
\end{table}
Both the current $J^5$ and $J^{05}$ contain a scalar (``good") diquark.
If the scalar diquark correlation is important in the correlators for
both currents, we expect to see similar behavior in both correlators.
The currents $J^{I}$ and $J^{3}$ contain a pseudoscalar diquark and an 
axial vector (``bad") diquark respectively.

Under exchanges of all indices for space, Dirac, flavor and color,
diquark structures should be antisymmetric. Since we are considering two quarks
in the color anti-triplet representation, they are antisymmetric with
respect to the color index. As for the space index, they are symmetric.
Thus the symmetric property of $C\Gamma$ ($\Gamma=\gamma_5$,
$\gamma_0\gamma_5$, $I$ and $\gamma_3$) determines the symmetric properties of
the diquarks under the exchange of the flavor index. For
$\Gamma=\gamma_5$, $\gamma_0\gamma_5$ and $I$, the diquark states are
antisymmetric under the exchange of Dirac indices. Therefore they have
to be
flavor isoscalars. For $\Gamma=\gamma_3$, the diquark state is a
flavor isovector.

The two point correlator for a current $J$ is defined as
$\langle\Omega|TJ(x)\bar{J}(0)|\Omega\rangle$,
where $|\Omega\rangle$ is the vacuum and $T$ is the time order operator.
For free massless quarks, the Euclidean coordinate space quark propagator  takes
the form 
\begin{equation}
\langle0|Tq(x)\bar{q}(0)|0\rangle=\frac{1}{2\pi^2}\frac{x_\mu\gamma_\mu}{x^4}.
\end{equation}
The index  $\mu$ is summed over.
For the current $J^5$ in Table~\ref{currenttab}, we have
\begin{equation}
\langle0|TJ^5(x)\bar{J}^5(0)|0\rangle=-\frac{15}{4\pi^6}
\frac{x_\mu\gamma_\mu}{x^{10}}.
\label{j5free}
\end{equation}
Similarly, we can get the free correlators for the other currents in
Table~\ref{currenttab}. They are the same as the result in
Eq.~(\ref{j5free}) except for a sign flip for $J^{05}$ and $J^I$
(for $J^3(J^{05})$, $x_3(x_0)=0$ is needed to get the same result). 
As was done in
Refs.~\cite{Chu:1993cn, Schafer:1993ra}, to reduce the sensitivity
on quark masses of the free correlators, which we will use to normalize
the interacting correlators,
it is convenient to multiply the correlators
with
$x_\nu\gamma_\nu$ and take the trace in the Dirac indices
for all the currents. For example, from Eq.(\ref{j5free})
we find
$R_0(x)\equiv\frac{1}{4}\Tr[\langle0|TJ^5(x)\bar{J}^5(0)|0\rangle
x_\nu\gamma_\nu]=-15/4\pi^6x^8$.
$R_0(x)$ is used to normalize the
interacting correlator $R(x)$, i.e. we will examine the ratio
$R(x)/R_0(x)$ for each current. In our lattice simulations, 
we use the free lattice correlators $R_0(x)$, rather than an analytic formula, to
do the normalization to reduce lattice artifacts.

Because of the
periodic boundary conditions in spatial directions, the correlators
receive contributions
from image points of the source, in addition to their ``direct"
contribution. When computing the free correlator, we should only keep
contributions, where
the source points for the three
quarks coincide, and so do the three sink points. They yield
the infinite volume correlator as we want.
The contributions from
the cross terms
are finite volume artifacts. In these cross terms, the three quark
sources sit on different image points and the three sink points
coincide. We expect that these cross terms affect the free and interacting
correlators differently.
For the
quenched system, the cross terms do not contribute since they
correspond to Polyakov loops encircling the entire lattice and
confinement forces their contributions to be zero. For the system with dynamical fermions,
propagation of different quarks to different image points involves intermediate states
in which the quarks separate a distance $L$ apart. Presumably, these contributions are
$\exp(-m_\pi L)$ effects. However for the free
theory, the cross terms give a nonzero contribution to the correlators,
which is power law in $L$, since
there is no confinement.

We cannot eliminate the image contributions from the free propagator. However, we can suppress them.
We do this via the method of
Ref~\cite{DeGrand:2001tm}: The free overlap fermion propagators are
calculated on a larger lattice ($24^4$) than the lattices for the interacting
simulations. 
Then the correlators in the
small volume are approximated as a sum of a
term from the source and a term from the nearest image point.

On the lattice, correlation functions in the coordinate space
for mesons, nucleon and $\Delta$ were calculated in
Refs.~\cite{DeGrand:2001tm, Chu:1993cn, Hands:1994cj}. In
\cite{Chu:1993cn, Hands:1994cj}, the correlators were fit to functions
from the corresponding parametrization of the spectral density functions
to an isolated resonance contribution and a continuum contribution.
Here we focus on the effects of diquark correlations on baryon correlators.
We do not try to do any fit of the correlators because our
lattice size is small.

Data at different values of $x$
are, of course, highly correlated, because they come from the same set of underlying configurations.
The statistical errors which we show in our plots are from a jackknife analysis.

\section{Numerical Results}\label{NumRes}
\begin{figure}
\begin{center}
\includegraphics[height=50mm,width=50mm]
{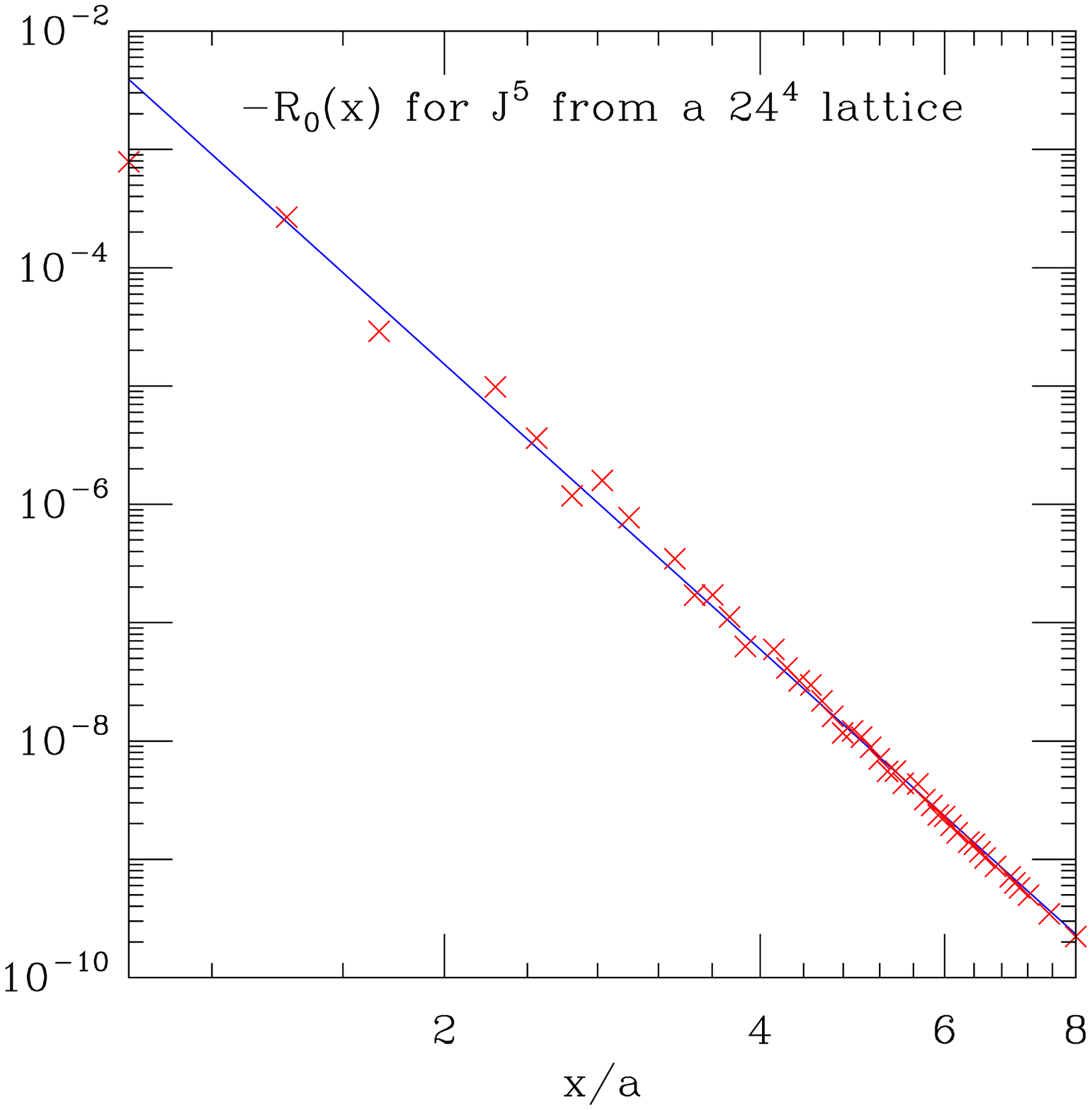}
\includegraphics[height=50mm,width=50mm]
{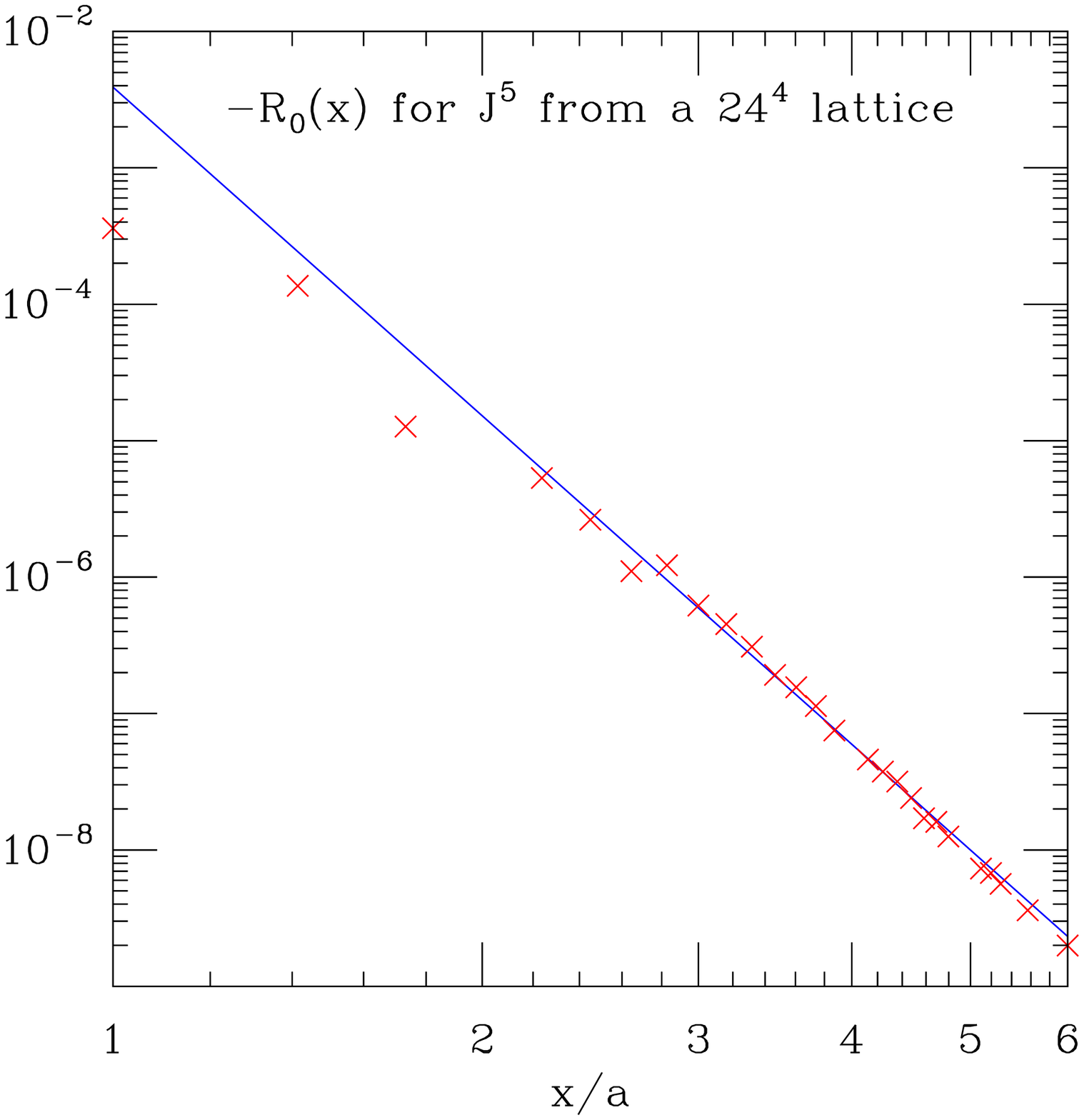}
\end{center}
\caption{Negative of the free correlation function $R_0(x)$
for $J^5$ from a $24^4$ lattice. 
The solid line is the continuum result
$15/4\pi^6x^8$. The left graph is for the lattice action used in the quenched simulation.
The right for the lattice action used in the dynamical simulation.
}
\label{free_r0x}
\end{figure}

Fig.~\ref{free_r0x} 
shows the continuum and lattice results for the free 
correlator $R_0(x)$ of the current $J^5$. We show one graph for 
each of the
quenched and dynamical simulations because the parameters for
the two ``kernel" Dirac operators of our overlap Dirac operator are slightly
different.
The lattice results agree quite well
with the continuum result except at very small distance, where lattice
discretization artifacts are apparent.
This behavior is seen for all the free correlators
we calculated.
In the following, we will restrict our discussion of the
results in the region $x/a<8$ (quenched) or $x/a<6$ (dynamical)
according to the lattice size.
We expect that the anisotropy introduced by the lattice will
cancel out to some extent in the ratio $R(x)/R_0(x)$.

\subsection{Quenched simulation results}
\begin{figure}
\begin{center}
\includegraphics[height=50mm,width=50mm]
{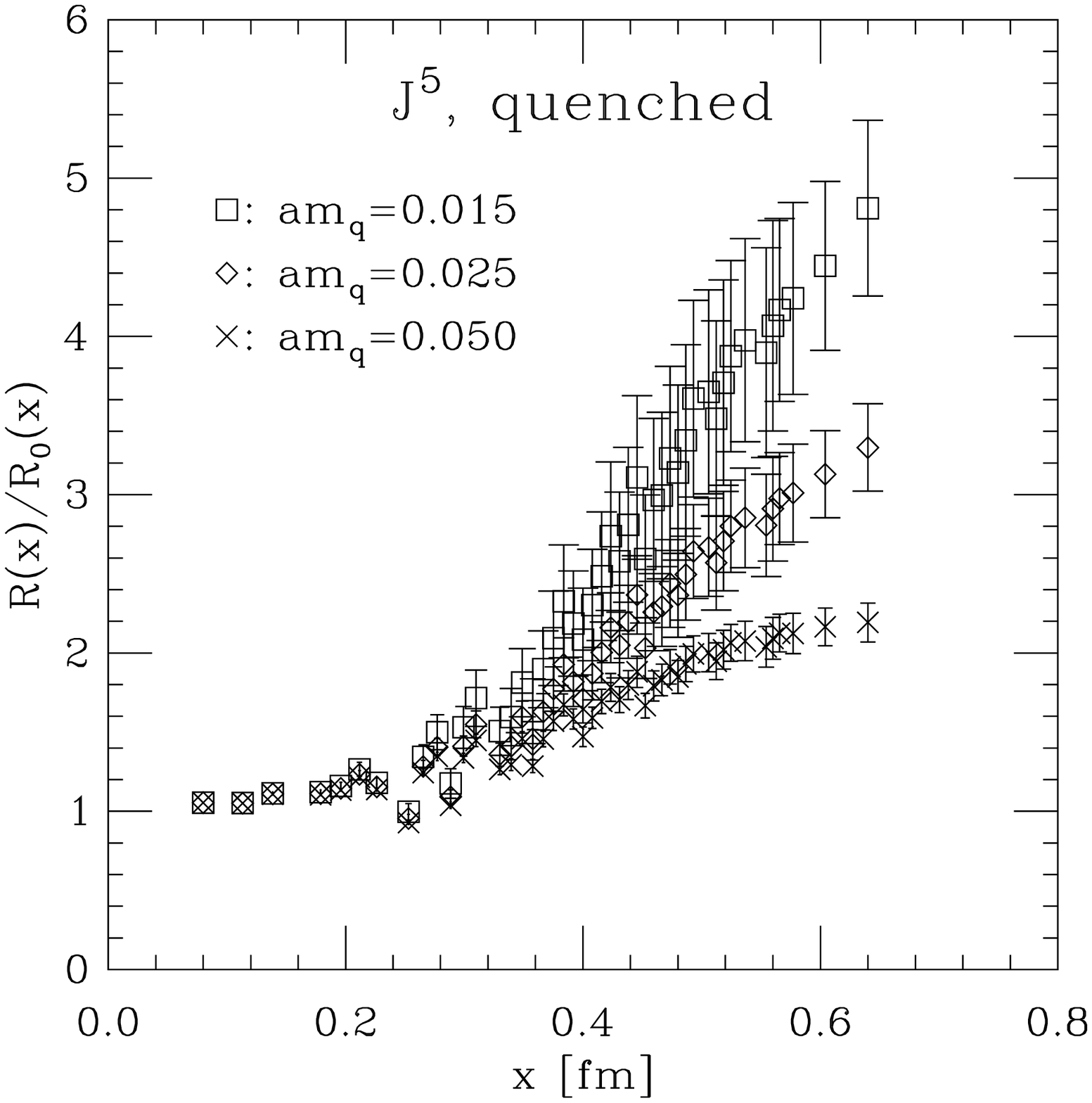}
\includegraphics[height=50mm,width=50mm]
{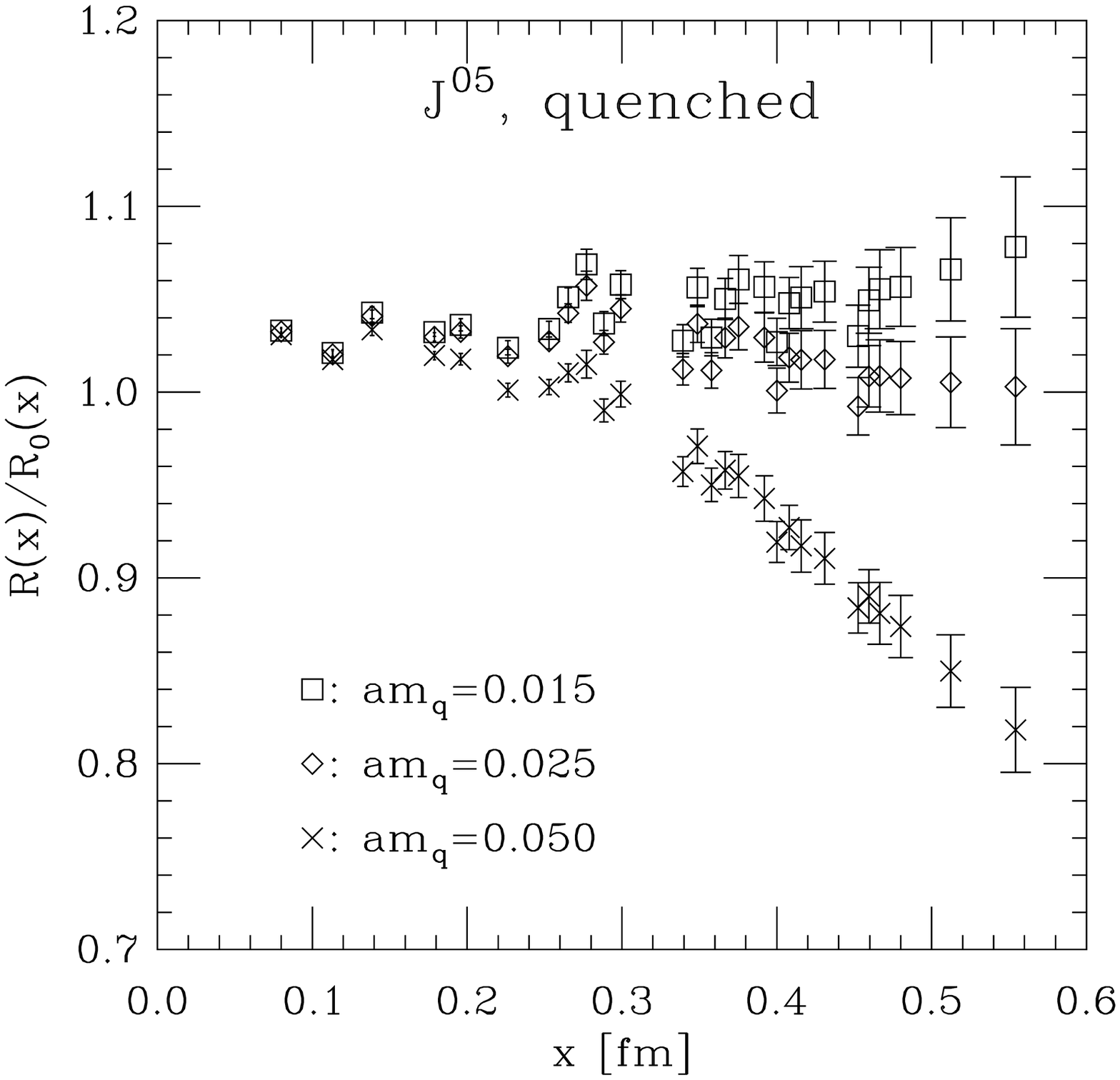}\\
\includegraphics[height=50mm,width=50mm]
{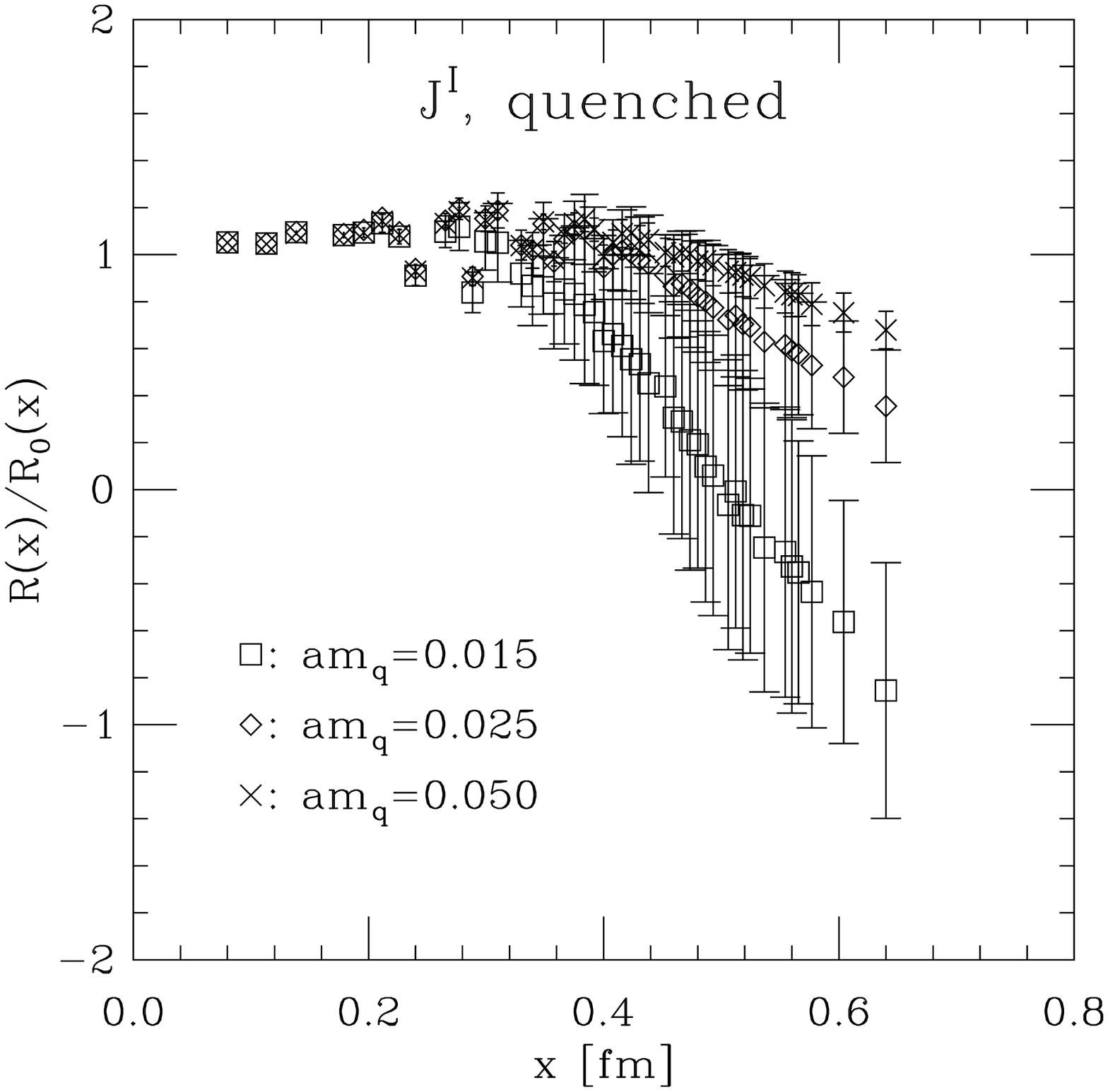}
\includegraphics[height=50mm,width=50mm]
{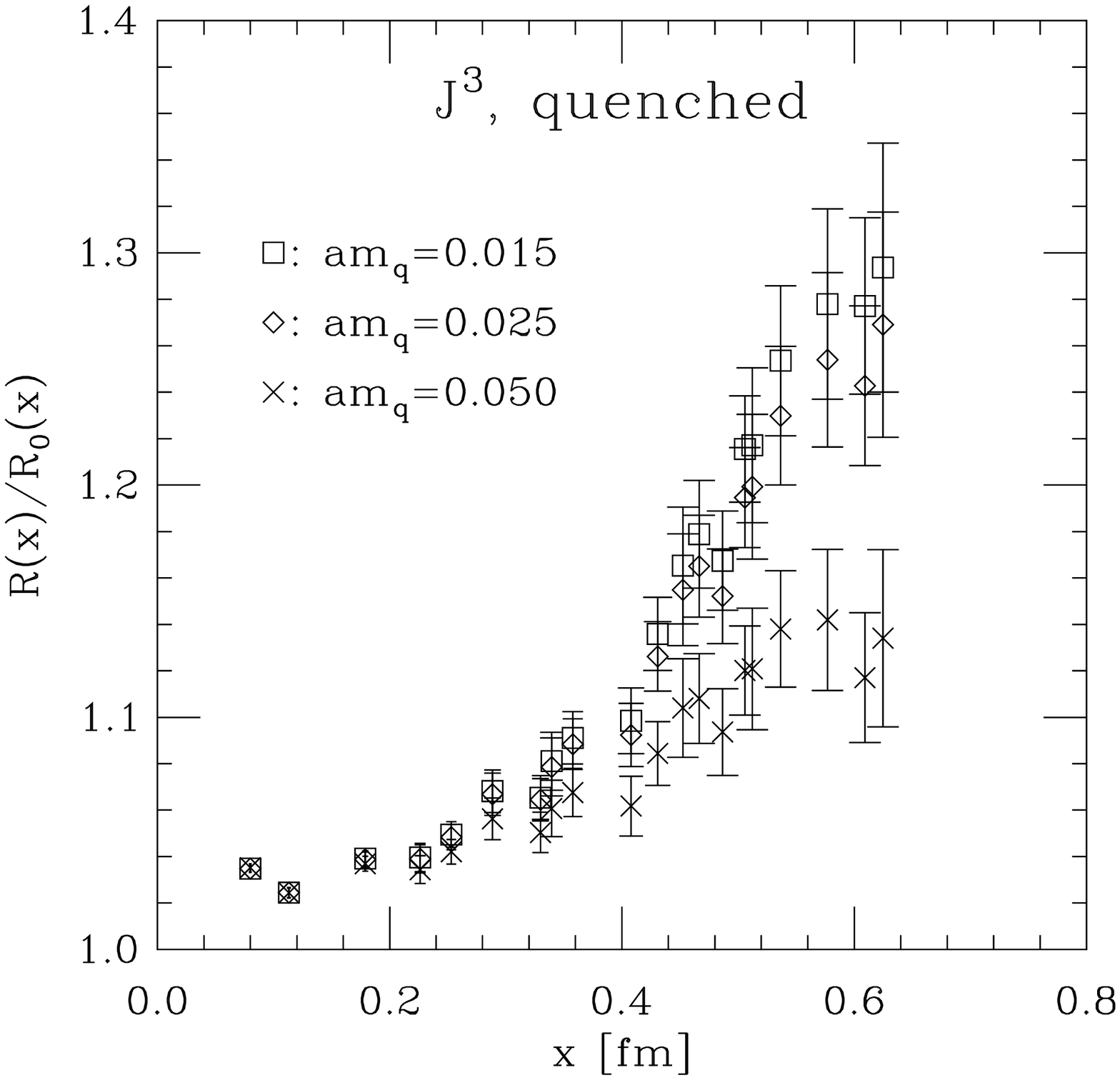}
\end{center}
\caption{Normalized correlation functions
$R(x)/R_0(x)$
for the four currents $J^5$, $J^{05}$, $J^I$ and $J^3$ from
the quenched simulations
for three quark
masses.}
\label{quen_all}
\end{figure}

Fig.~\ref{quen_all} shows the normalized correlation 
functions
$R(x)/R_0(x)$ against the physical distance $x$
for the four currents $J^5$, $J^{05}$,
$J^I$ 
and $J^3$ for three quark masses. The first graph shows that
the correlator for $J^5$ is attractive ($>1$) and
the attraction increases  as the quark mass decreases. 
For $J^{05}$,
at the largest quark mass the correlation function
is repulsive ($<1$), which is very different from the correlator for
$J^{5}$ although both of them contain a scalar diquark.
As the quark mass decreases, the correlator for $J^{05}$ becomes flat and
then seems to curve up, showing some attraction. 
For the current $J^I$, which contains a pseudoscalar diquark, we
find that at the smallest quark mass the normalized correlation function
tends to go negative at large distance.
This behavior is very
similar to what was  observed for the scalar meson in
Ref.~\cite{DeGrand:2001tm}. It was argued in Ref.~\cite{DeGrand:2001tm}
that zero modes are the sources of the negativity. In the following,
we will see zero modes do make a big difference here.
The last graph in Fig.~\ref{quen_all}
is for the current $J^3$, which contains
an axial vector (``bad") diquark.
Some attraction shows up above $x=0.32$
fm and seems to increase as the quark mass decreases.
Comparing with $J^5$,
the attraction in this channel is much
weaker. 

\begin{figure}
\begin{center}
\includegraphics[height=50mm,width=50mm]
{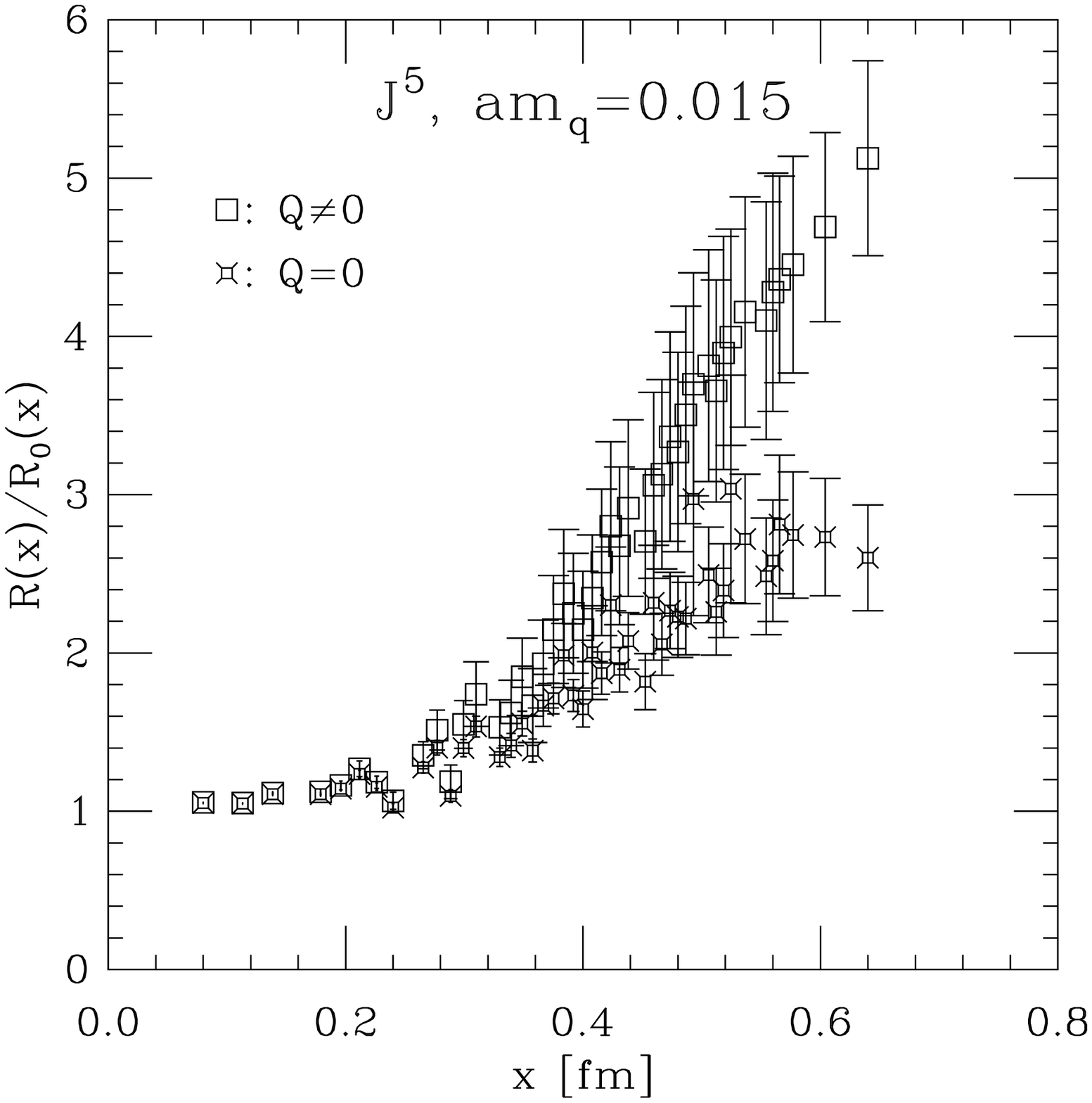}
\includegraphics[height=50mm,width=50mm]
{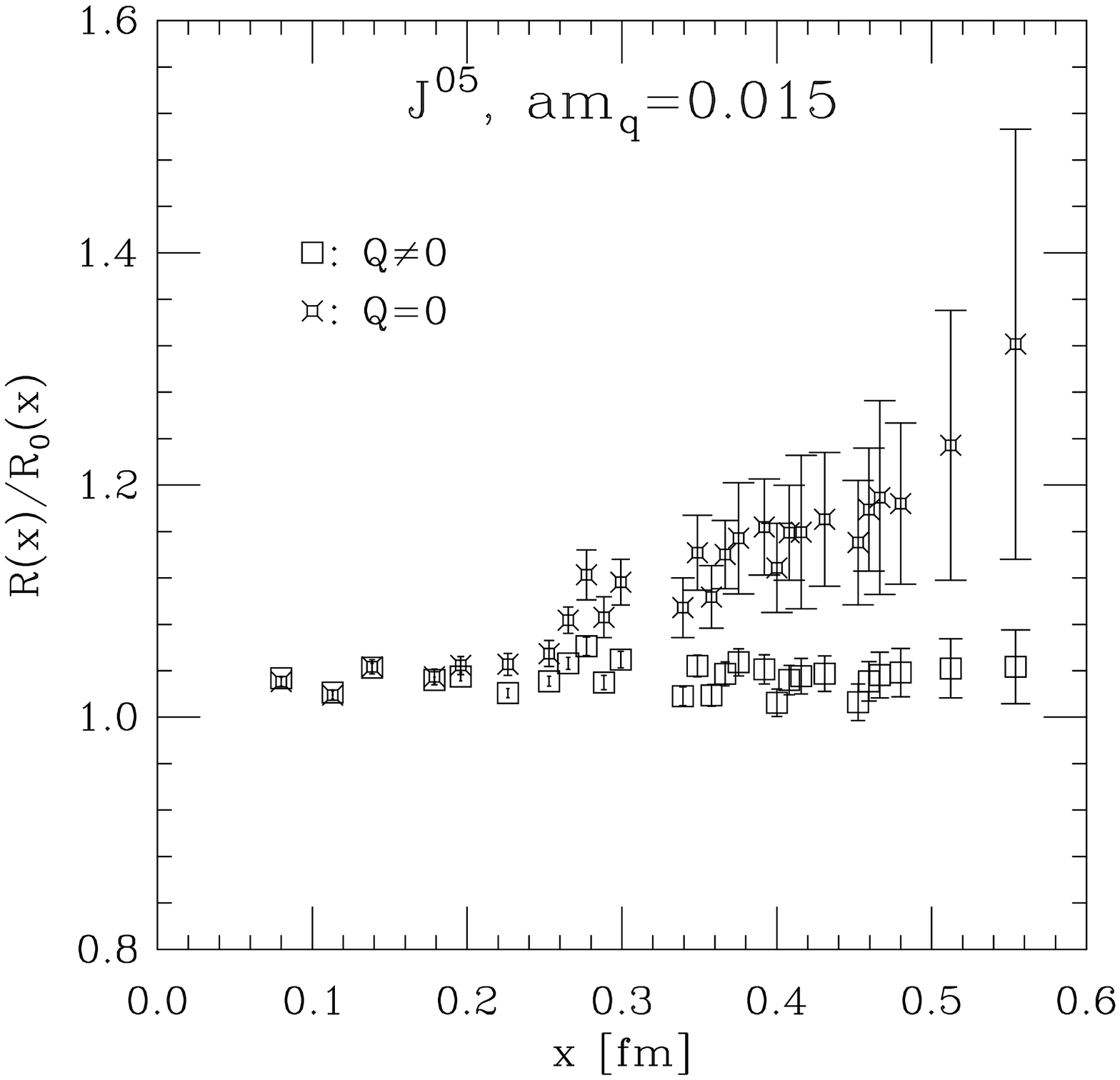}\\
\includegraphics[height=50mm,width=50mm]
{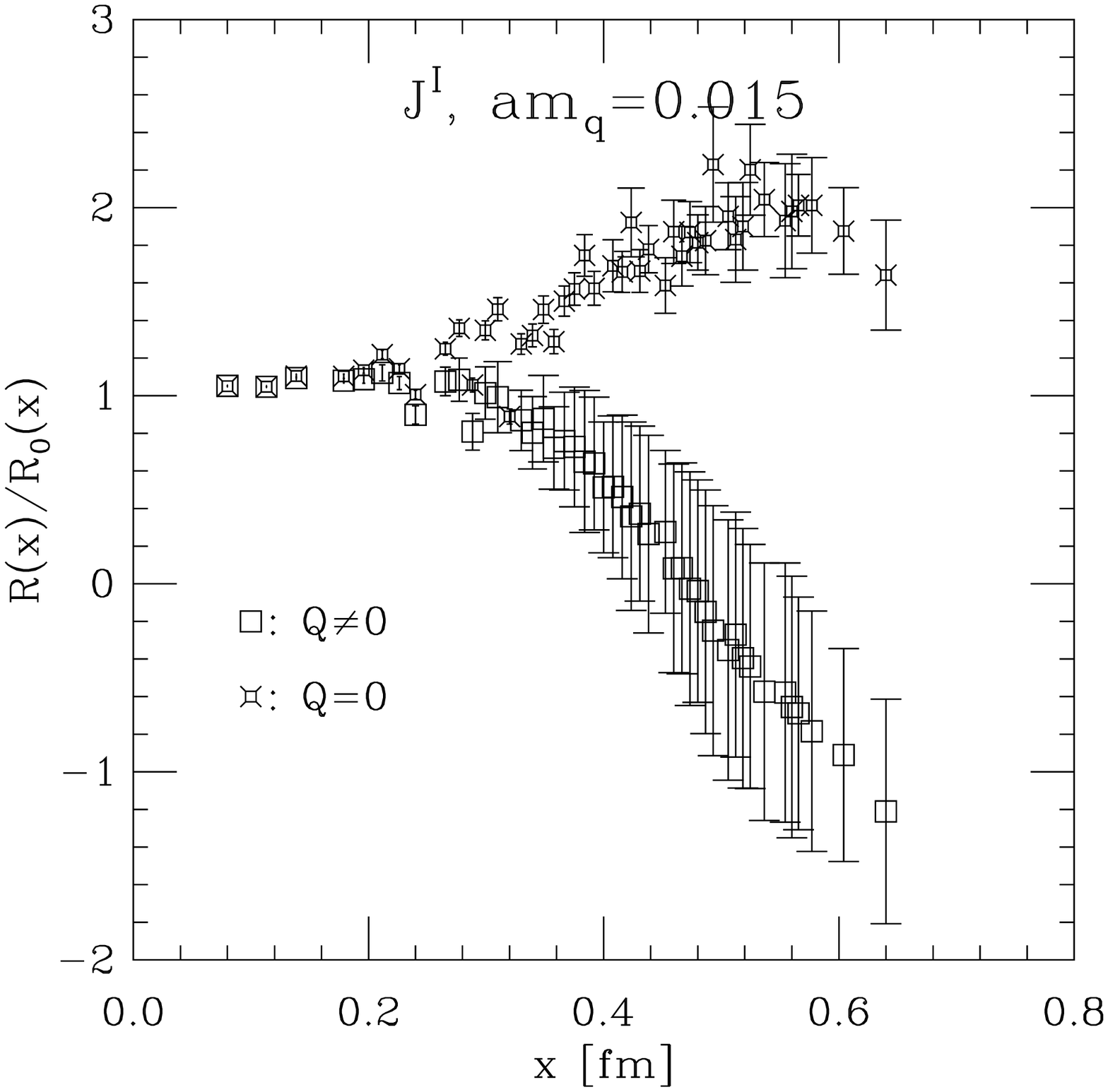}
\includegraphics[height=50mm,width=50mm]
{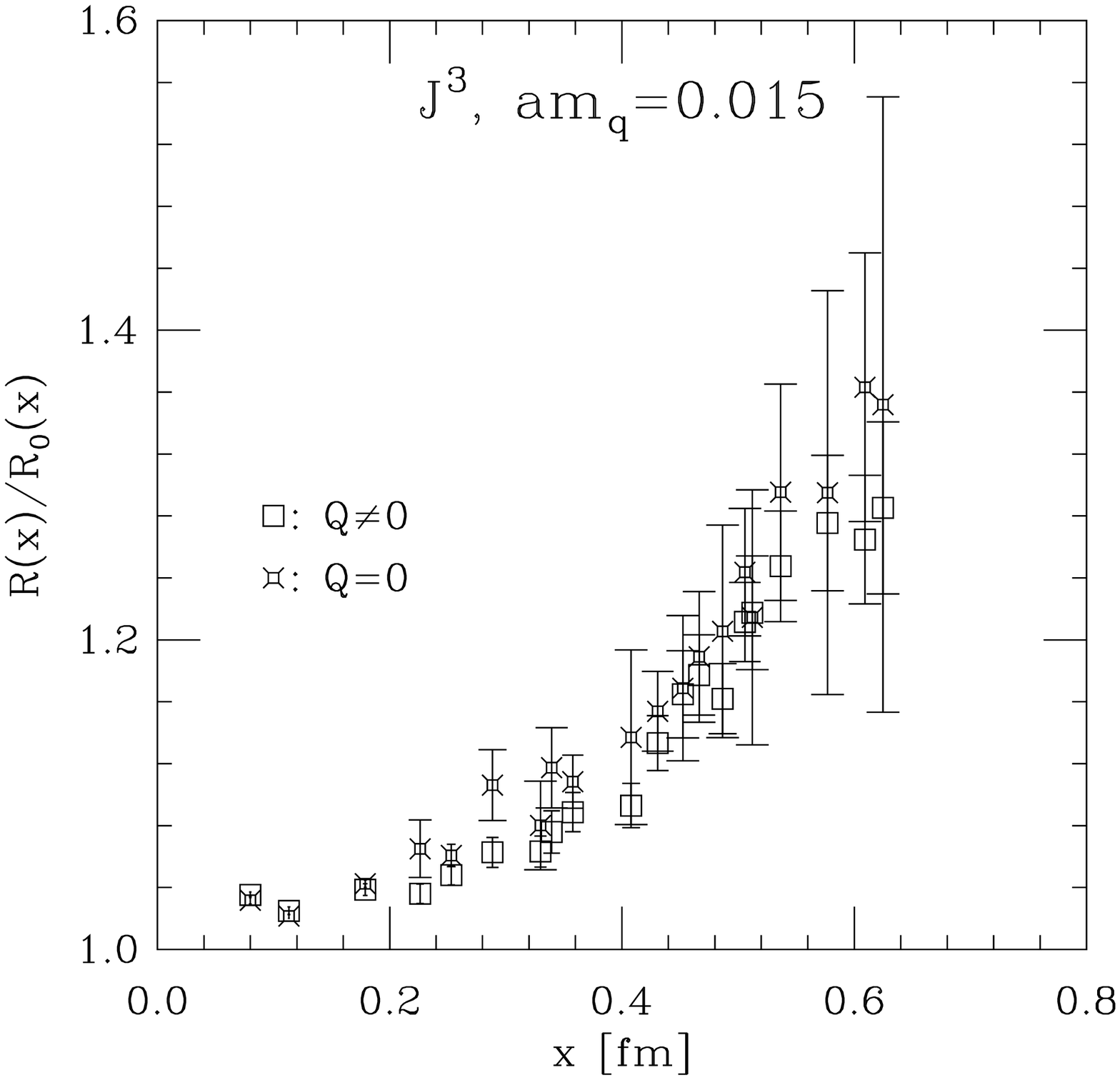}
\end{center}
\caption{The normalized correlation functions
$R(x)/R_0(x)$
for the four currents for the lowest quark mass, from quenched data.
We show the comparison of the correlators
from configurations
with $Q\neq0$ and
from configurations with $Q=0$.}
\label{quen_q0.015}
\end{figure}

\begin{figure}
\begin{center}
\includegraphics[height=50mm,width=50mm]
{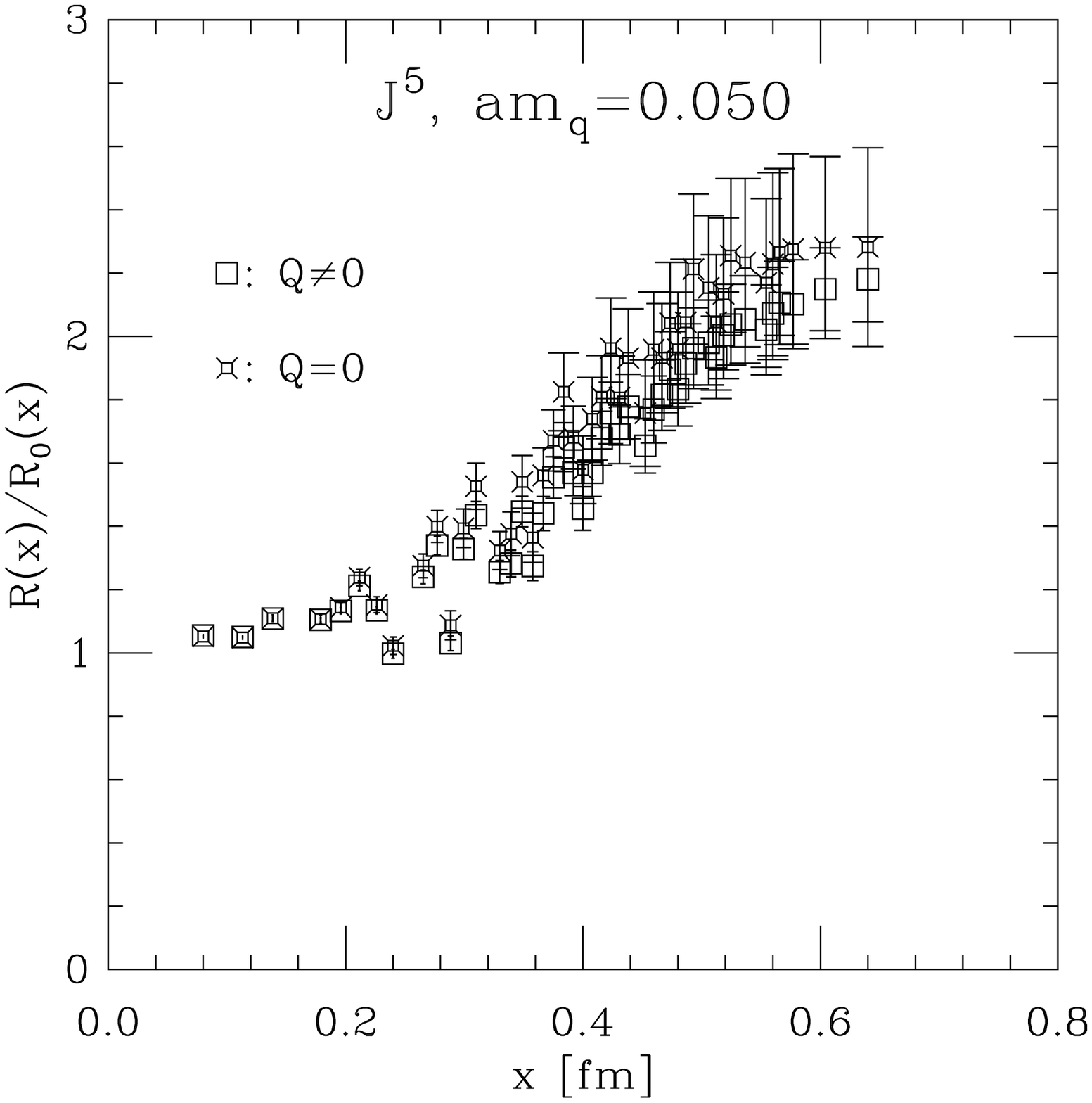}
\includegraphics[height=50mm,width=50mm]
{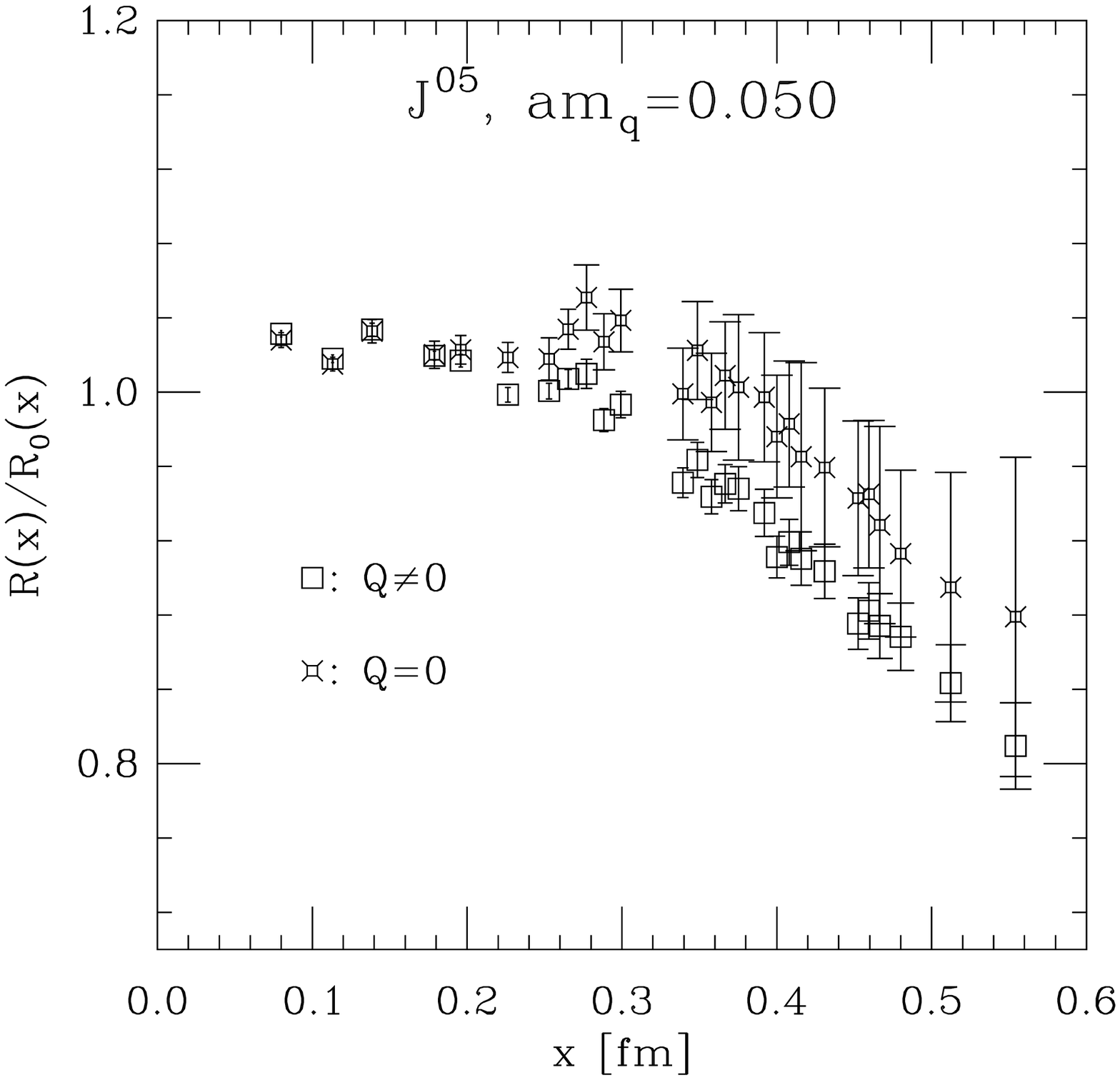}\\
\includegraphics[height=50mm,width=50mm]
{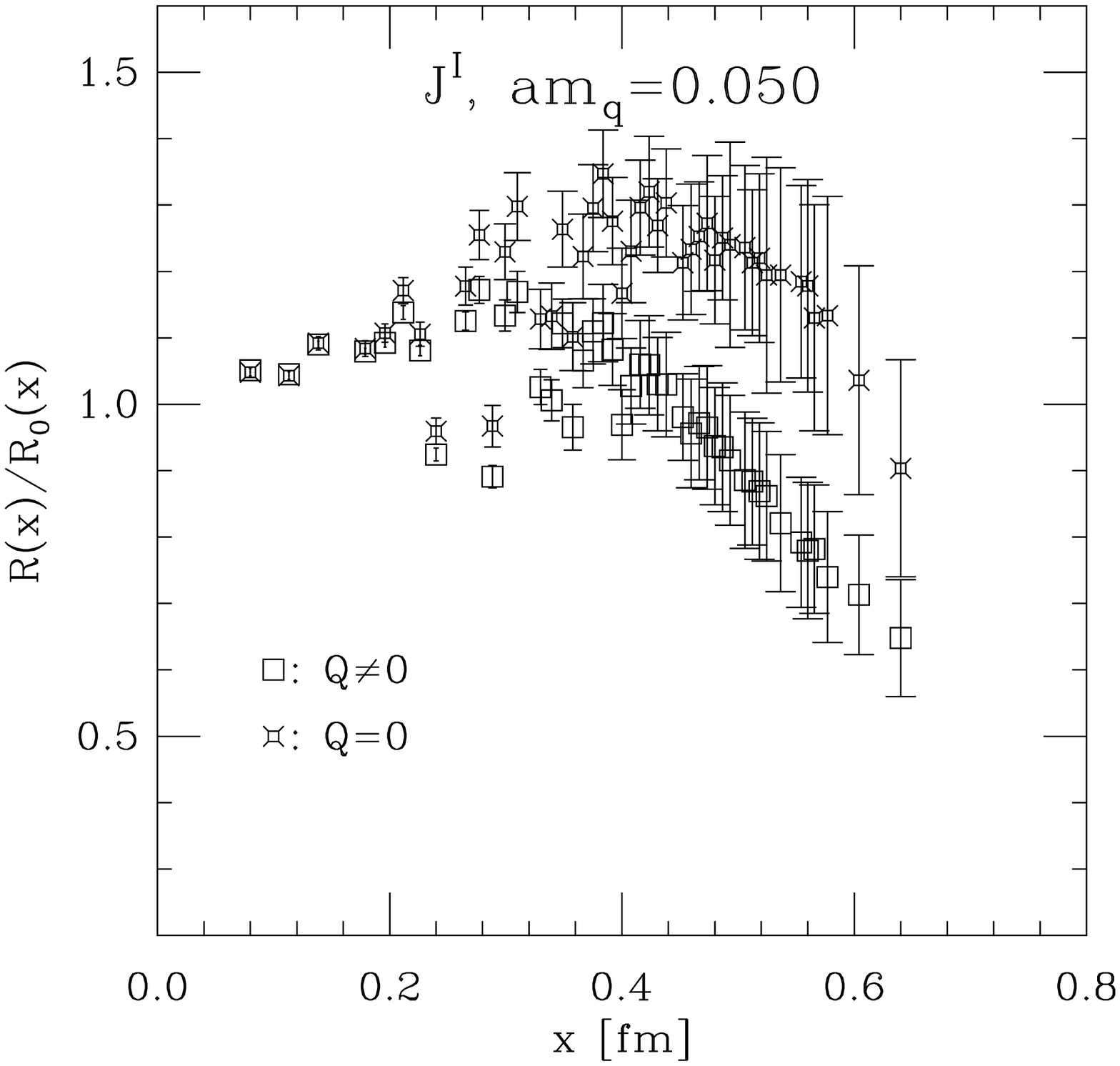}
\includegraphics[height=50mm,width=50mm]
{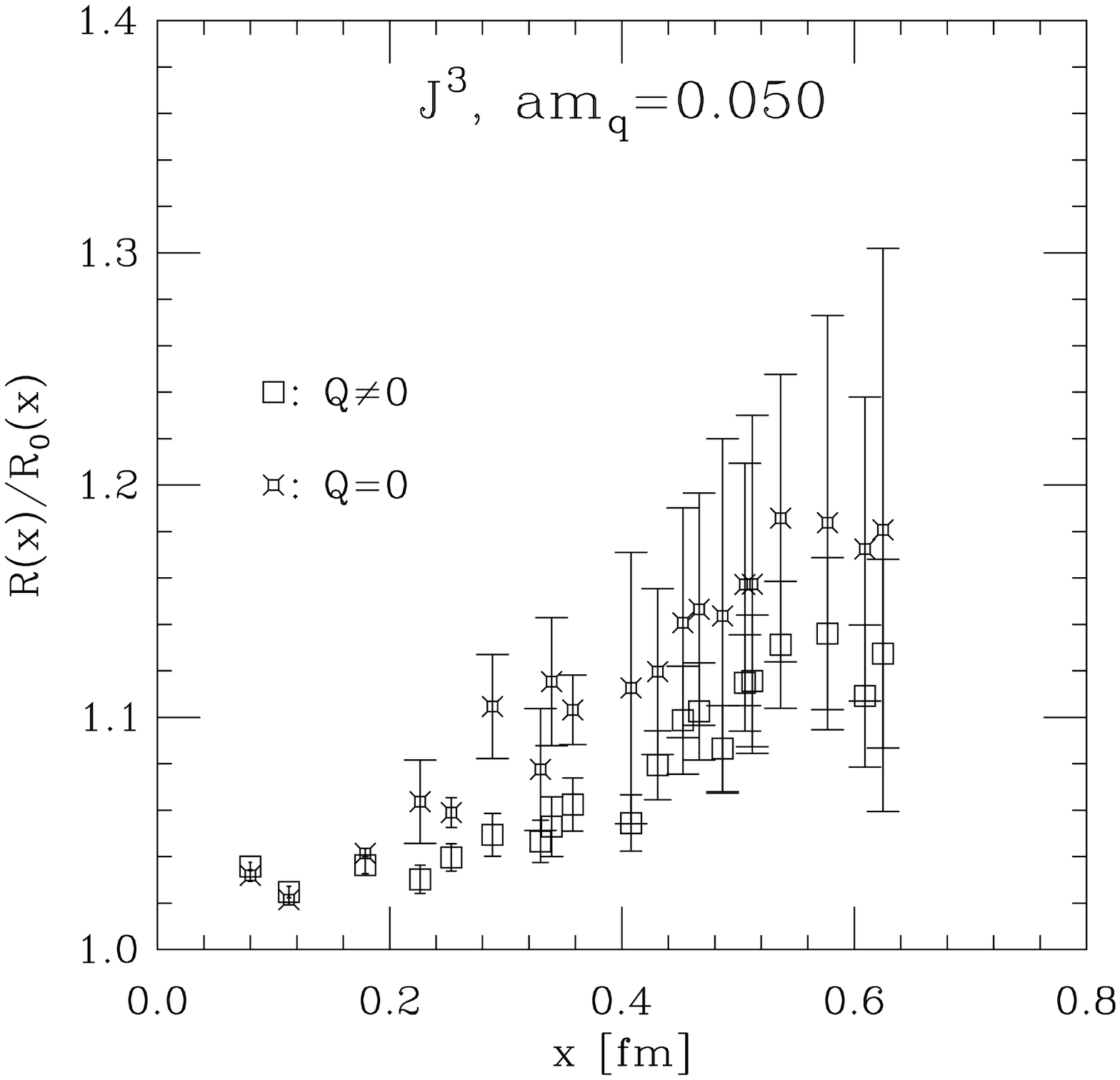} 
\end{center}
\caption{The normalized correlation functions
$R(x)/R_0(x)$
for the four currents for $am_q=0.050$, from quenched data.
We show the comparison of the correlators
from configurations
with $Q\neq0$ and
from configurations with $Q=0$.}
\label{quen_q0.050}
\end{figure}

To see zero mode effects, we separate the 40 configurations into two
groups: 35 of them with topological charge $Q\neq0$ and the other 5 with
$Q=0$. i.e. in the first group the quark propagator 
contains zero modes while in the second one
it does not. From each group we compute the correlation functions
and compare. The
results are shown in Fig.~\ref{quen_q0.015} for our smallest quark mass
and Fig.~\ref{quen_q0.050} for the largest quark mass.
For the current $J^5$, at the lowest quark mass
the attraction at large $x$ ($x>0.4$ fm) is larger in the  $Q\ne 0$ sector.
For $J^{05}$, the zero mode effects
are in the opposite direction of those for $J^{5}$.
The correlator from
the $Q=0$ group (no zero mode) is greater than the correlator from the
$Q\neq0$ group. i.e., zero modes make the $J^{05}$ correlator less attractive.
For $J^{I}$, the zero modes contribute strongly to the repulsion 
seen at the lowest quark
mass 0.015 so
that $R(x)/R_0(x)$ becomes negative at large $x$. 
For $J^{3}$, when we compare the correlators from the $Q\neq0$ group and from
the $Q=0$ group, we find no difference within error bars.
Zero mode effects are not important in this channel.

In Fig.~\ref{quen_q0.050}, we see similar, but smaller, zero mode effects in each channel
for $am_q= 0.050$. This is natural given how zero modes scale with quark mass.

\begin{figure}
\begin{center}
\includegraphics[height=50mm,width=50mm]
{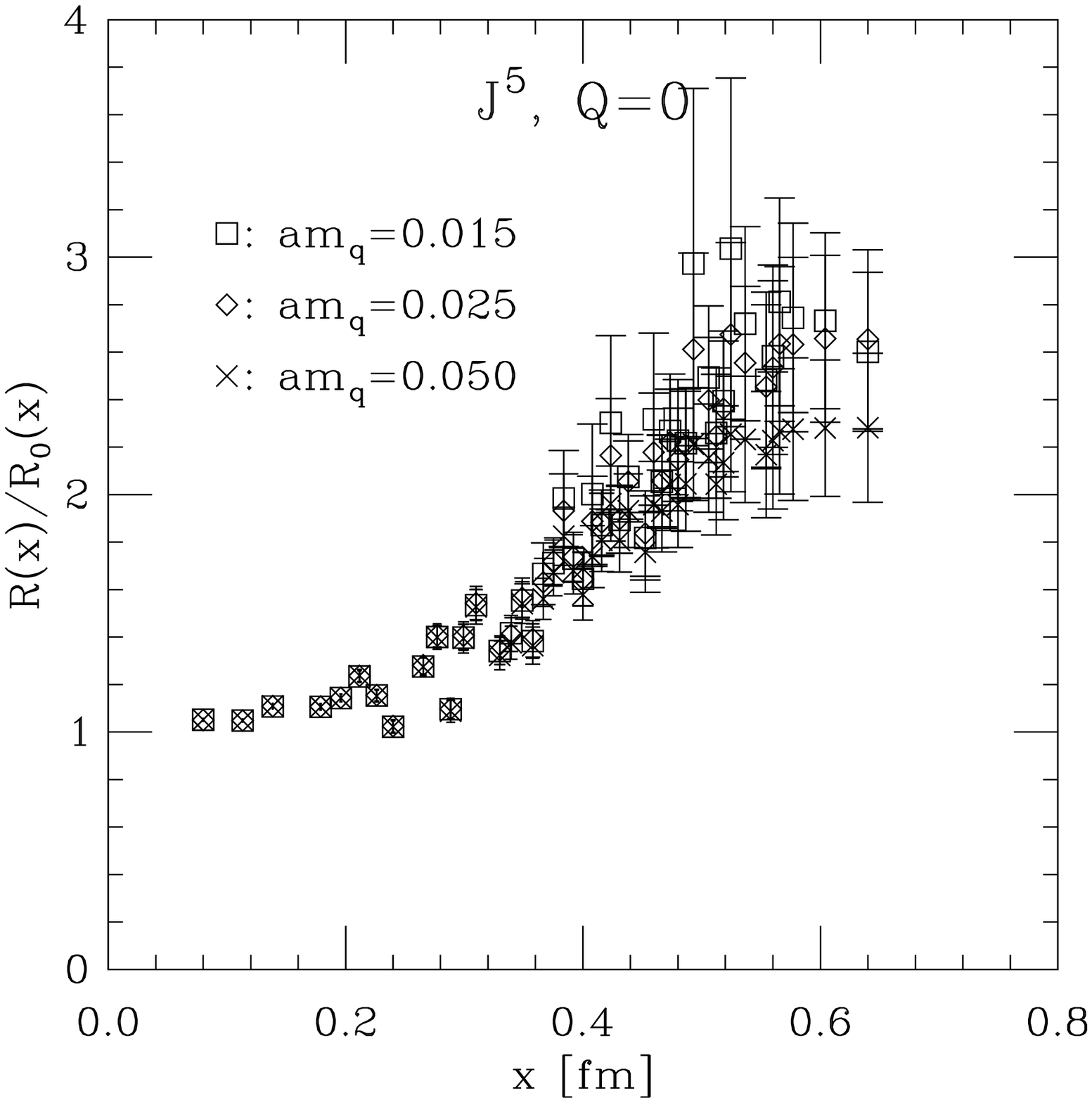}
\includegraphics[height=50mm,width=50mm]
{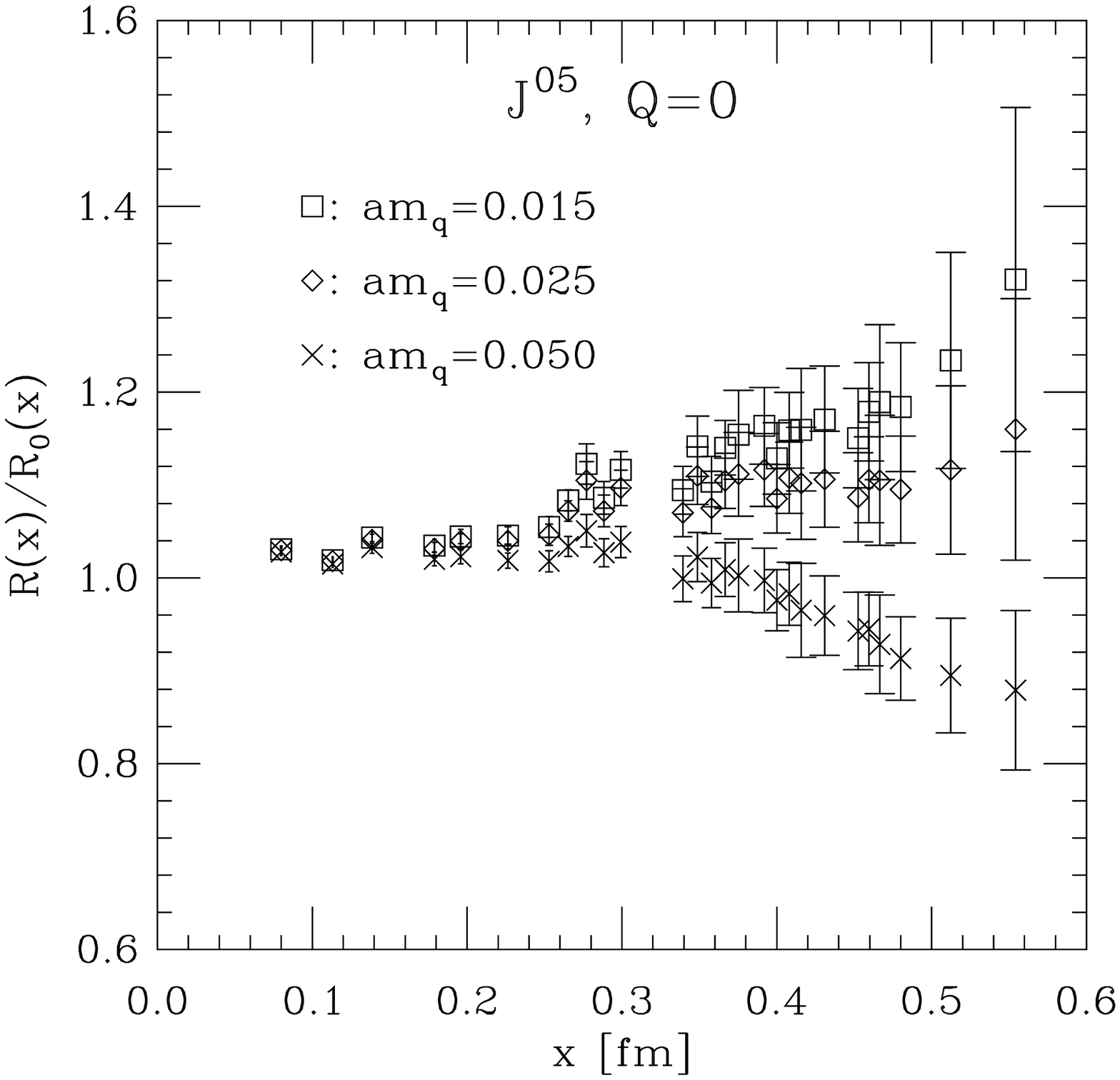}\\
\includegraphics[height=50mm,width=50mm]
{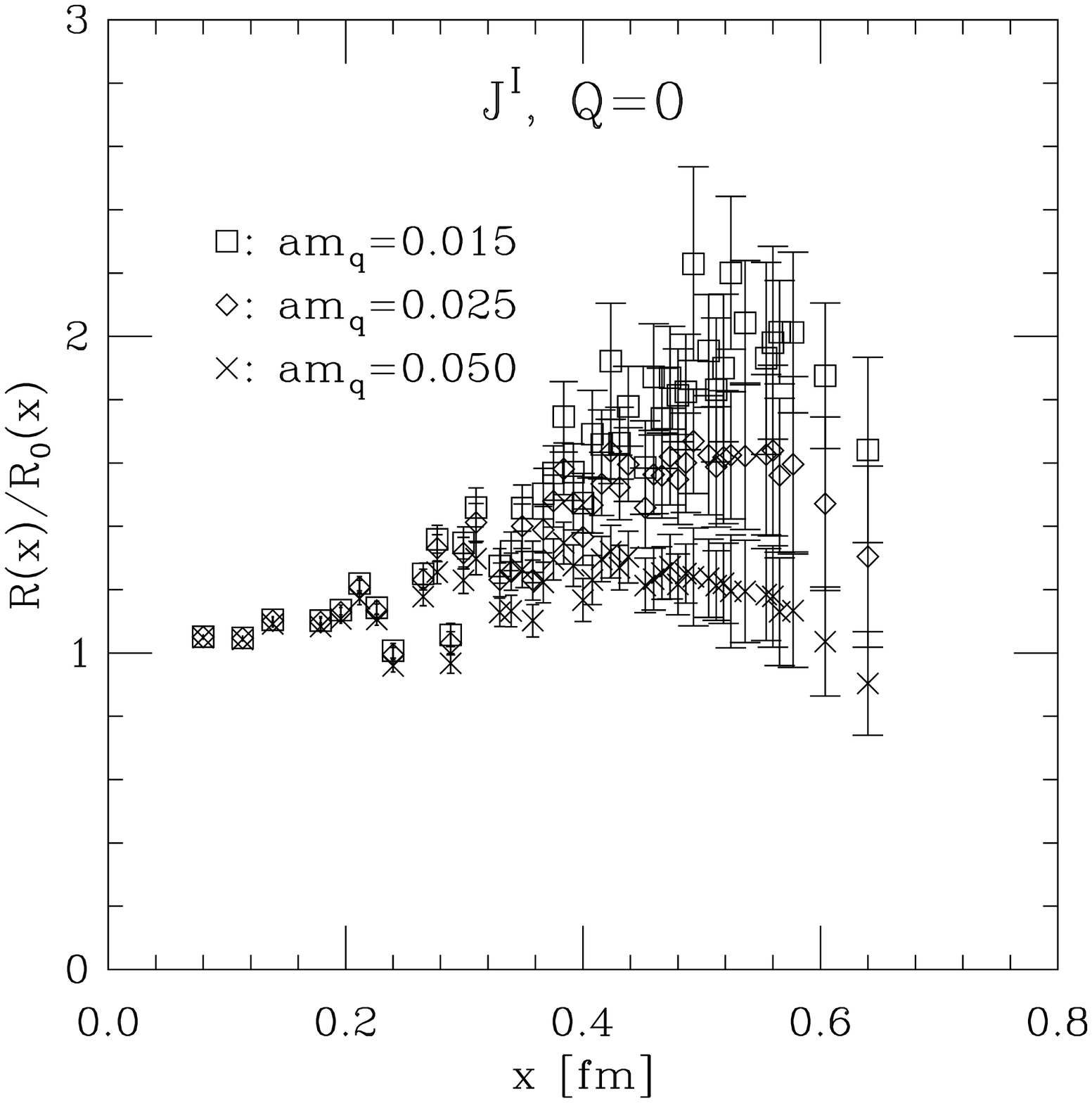}
\includegraphics[height=50mm,width=50mm]
{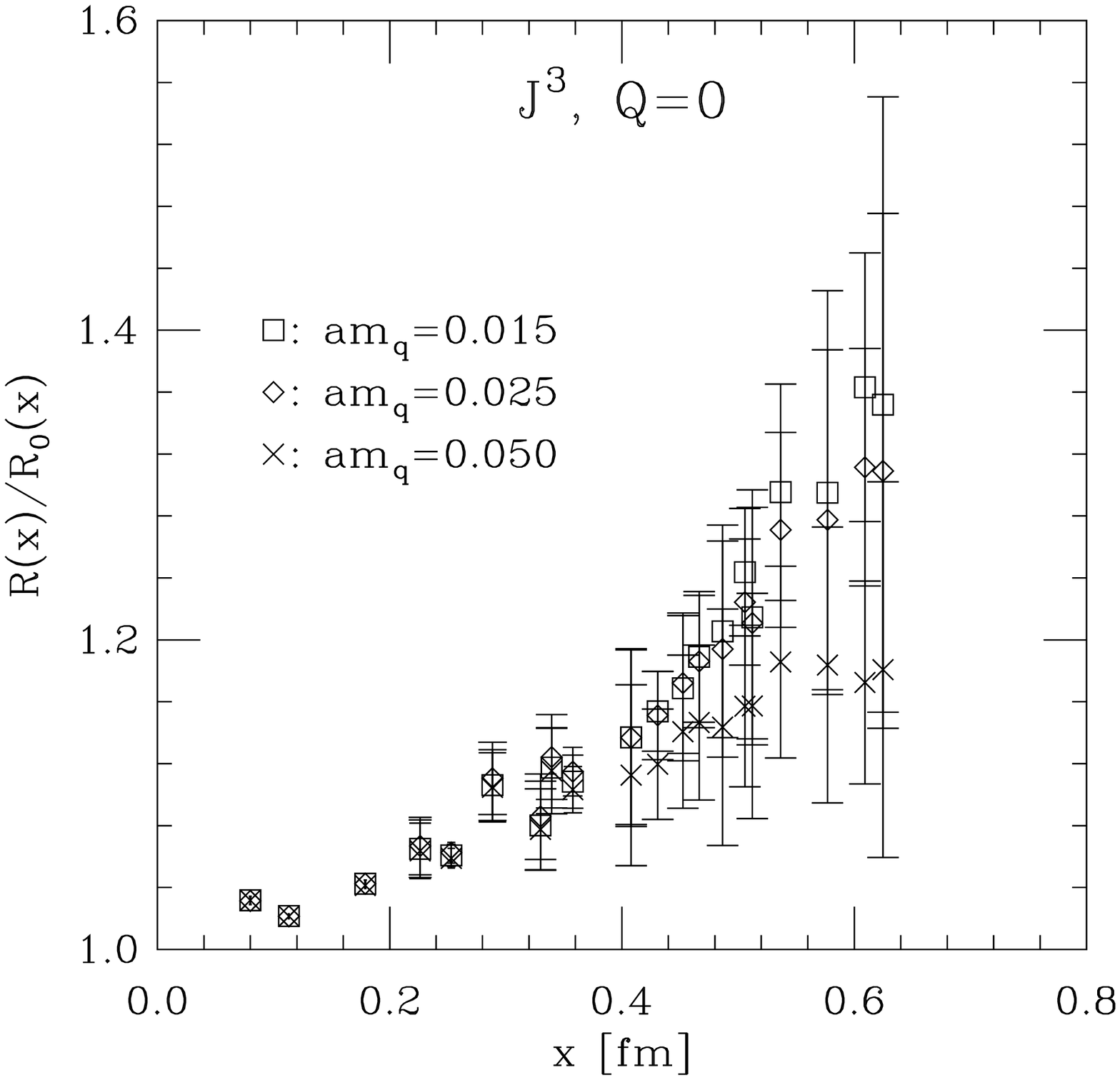}
\end{center}
\caption{The normalized correlation functions
$R(x)/R_0(x)$
for the four currents computed from
the configurations
with $Q=0$ for the three quark
masses.}
\label{quen_all_q0}
\end{figure}

If we only use the   $Q=0$ configurations from our quenched simulation,
we obtain  the correlation functions  shown in
Fig.~\ref{quen_all_q0}. 
Comparing with Fig.~\ref{quen_all}, we see that
in the $J^5$ channel the attraction is weaker and its dependence on the
quark mass is also weaker. The $J^{05}$ channel shows a little bit
more attraction than in Fig.~\ref{quen_all} at the smallest quark 
mass $am_q=0.015$. The biggest change is in the $J^I$ channel. The $J^I$ correlators
for all three quark masses are attractive in Fig.~\ref{quen_all_q0}.
In contrast,  in Fig.~\ref{quen_all}, they are repulsive and even negative.
For $J^3$, there is no difference within error bars between 
Fig.~\ref{quen_all} and Fig.~\ref{quen_all_q0} since zero mode effects
are small in this channel.

\begin{figure}
\begin{center}
\includegraphics[height=50mm,width=50mm]
{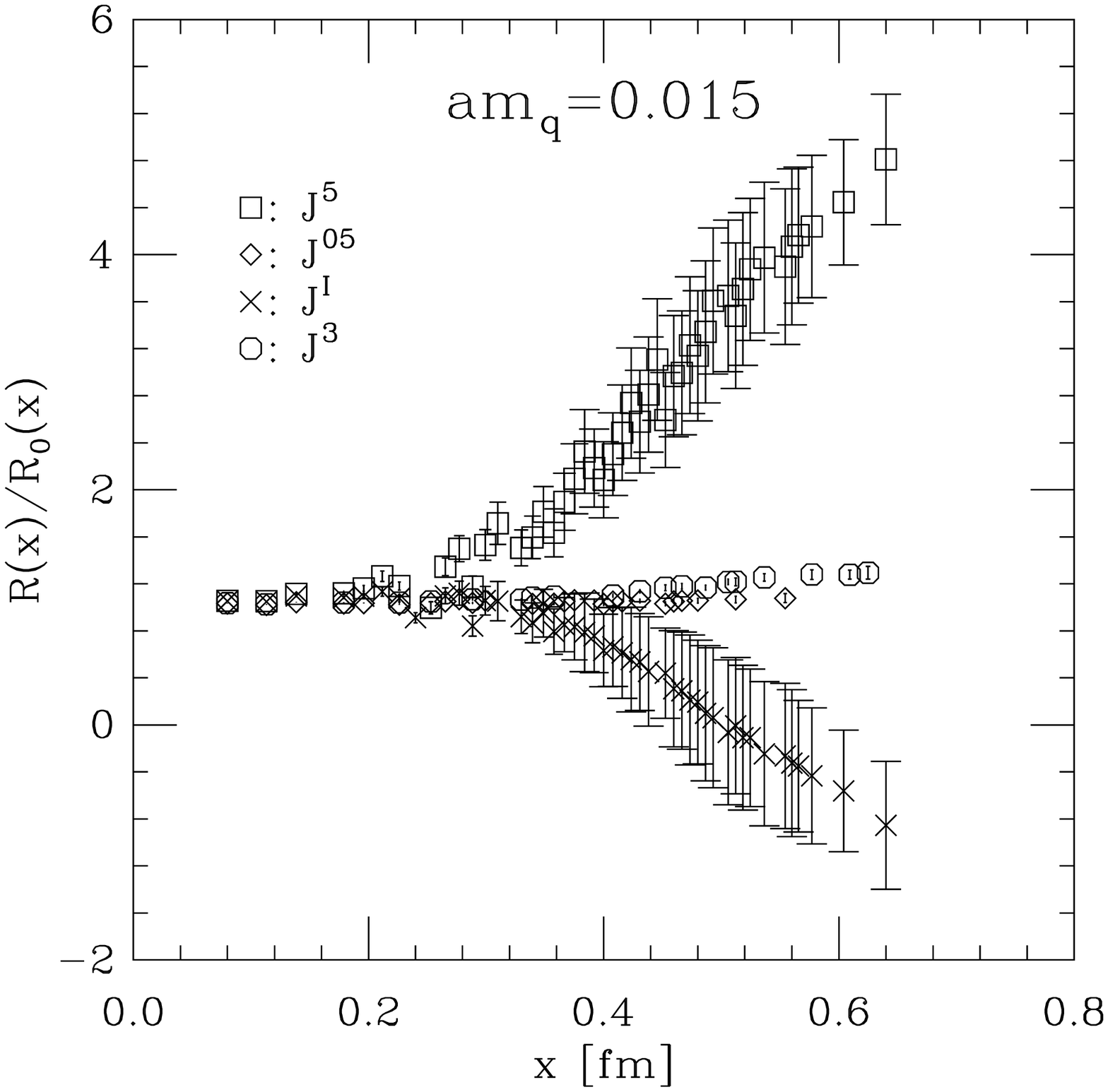}
\includegraphics[height=50mm,width=50mm]
{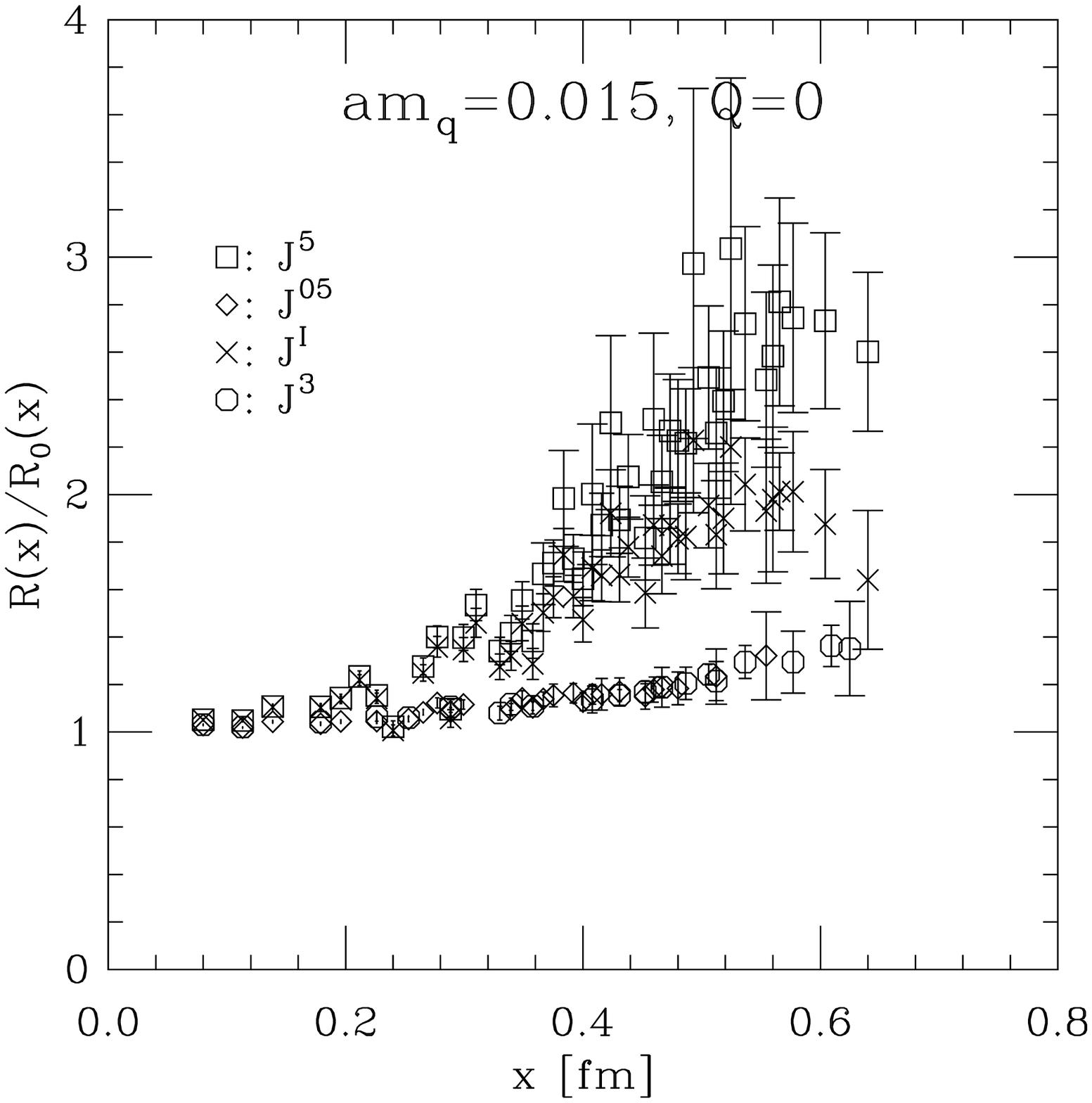}
\end{center}
\caption{Comparison of the four normalized correlators
$R(x)/R_0(x)$ from the quenched simulation for the lowest quark mass.
The left graph is from all configurations. The right from
the configurations with $Q=0$.}
\label{quen0.015}
\end{figure}
To summarize the quenched results, we find that the attraction in
the $J^5$ channel, which contains a scalar diquark
structure, is the strongest. The $J^I$ channel is repulsive and even becomes negative
at large distance as the quark mass decreases.
The $J^{05}$ channel is repulsive at the large quark mass, but
turns into attractive at small quark masses. 
Its behavior is quite different from the correlator for $J^5$
even though $J^{05}$
also contains a scalar diquark.
The axial vector channel is weakly attractive and the attraction
seems to increase when we go to light quark masses. 
The comparison of the four correlators obtained from all 40 configurations
at the lowest quark mass $am_q=0.015$
is shown in the left graph in Fig.~\ref{quen0.015}. The $J^5$
channel is very different from the other channels.

By 
comparing the correlators from different topological groups,
we find that at small quark masses
both the attraction in the scalar diquark channel ($J^5$) and the
repulsion in the pseudoscalar diquark channel ($J^I$) have big
contributions from configurations with zero modes.
In the $J^{05}$ channel, the zero
modes make the correlator a little less attractive.
In the right graph of Fig.~\ref{quen0.015}, we show the
comparison of the four correlators obtained from configurations
without zero modes
at the lowest quark mass 0.015.
The difference between channel $J^5$ and $J^I$ shrinks considerably.
The attractions in these two channels are similar and
stronger than the attractions in $J^{05}$ and $J^3$. 

In full QCD simulations, configurations with high $Q$ values 
are suppressed by the
fermion determinant at small quark masses. But certainly not
all configurations are with $Q=0$.
Thus we expect to see a picture somewhere between 
the two
graphs in Fig.~\ref{quen0.015} at small quark masses in our
dynamical simulation.

\subsection{Dynamical simulation results}

\begin{figure}
\begin{center}
\includegraphics[height=50mm,width=50mm]
{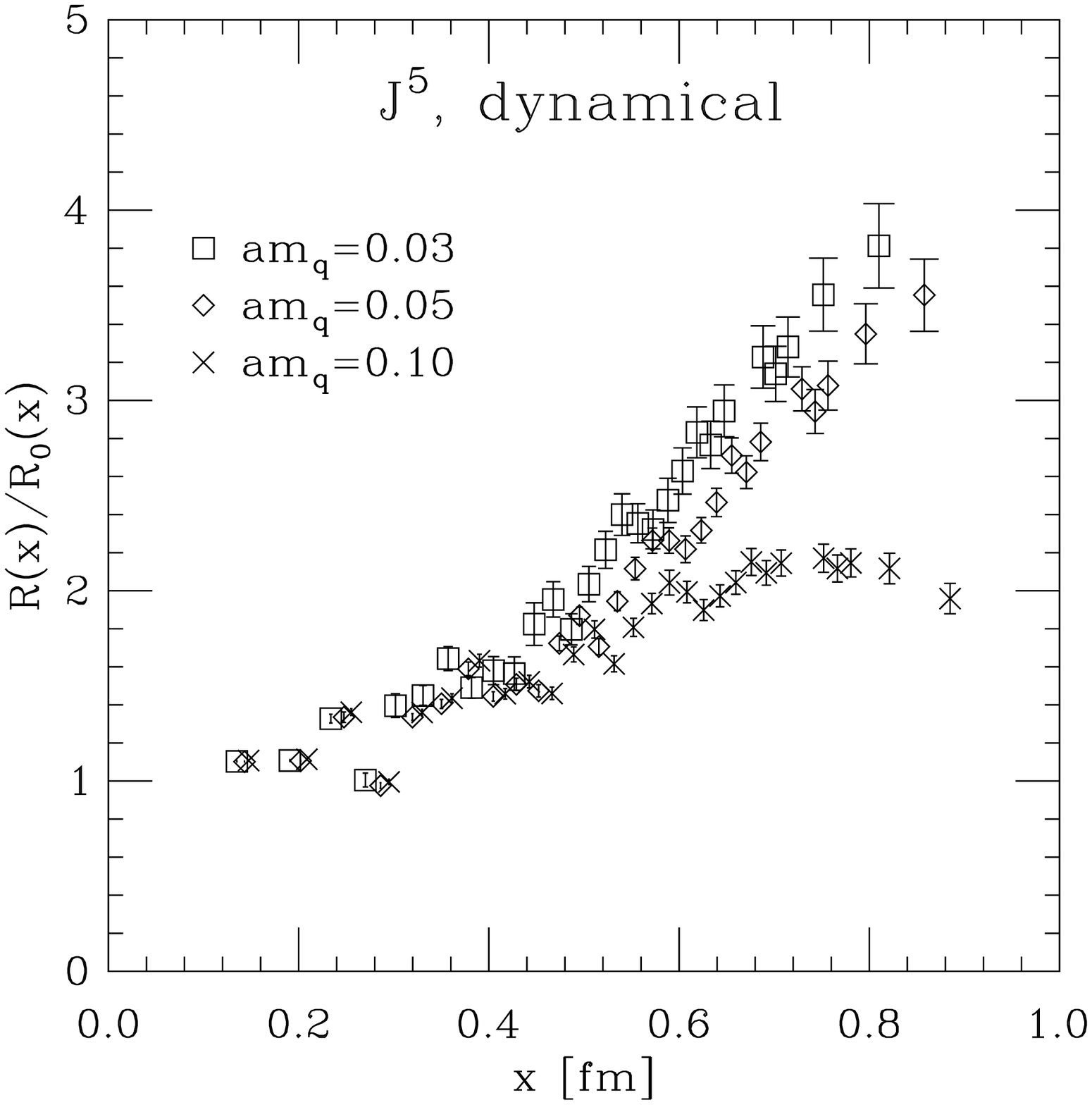}
\includegraphics[height=50mm,width=50mm]
{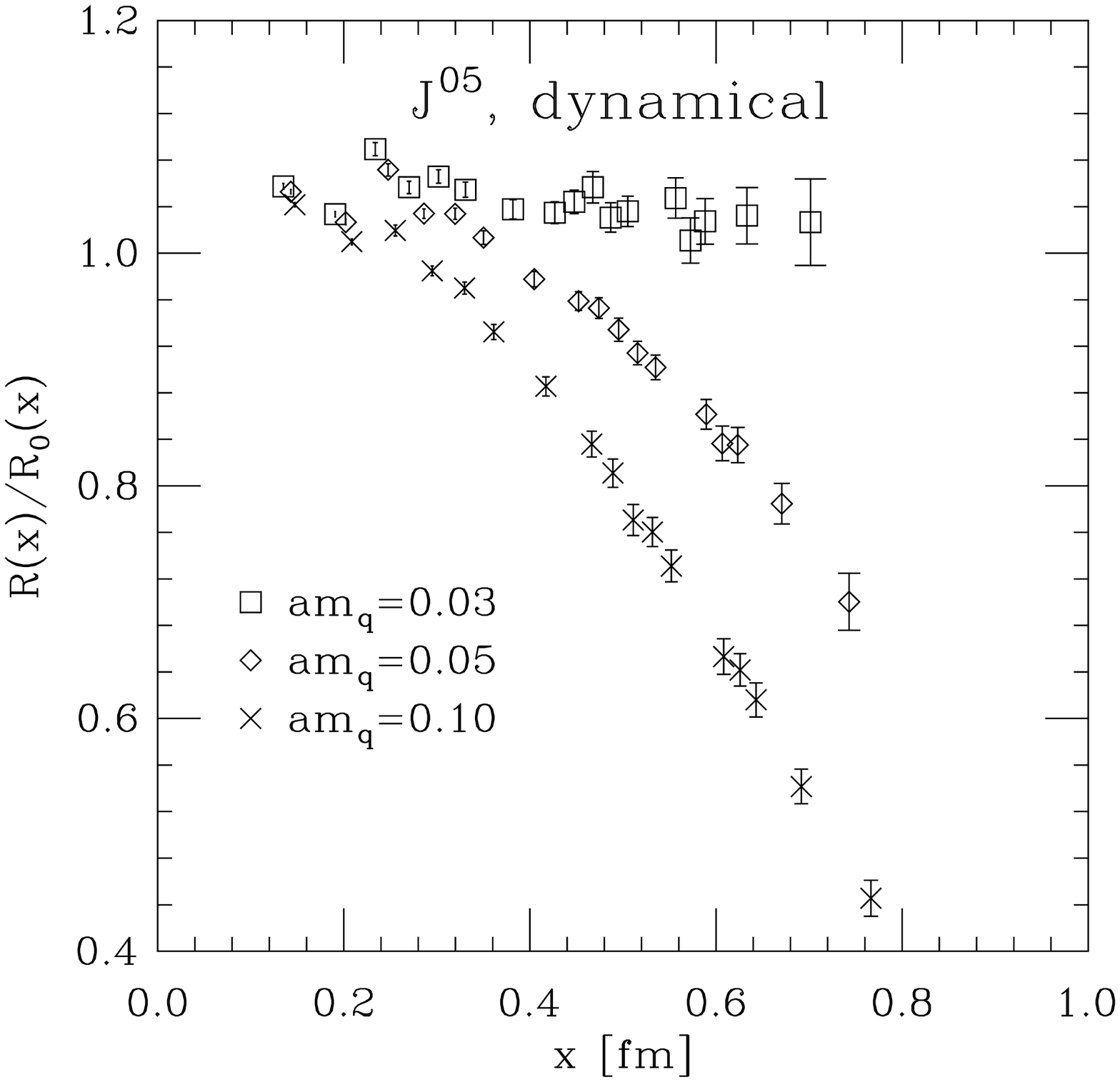}\\
\includegraphics[height=50mm,width=50mm]
{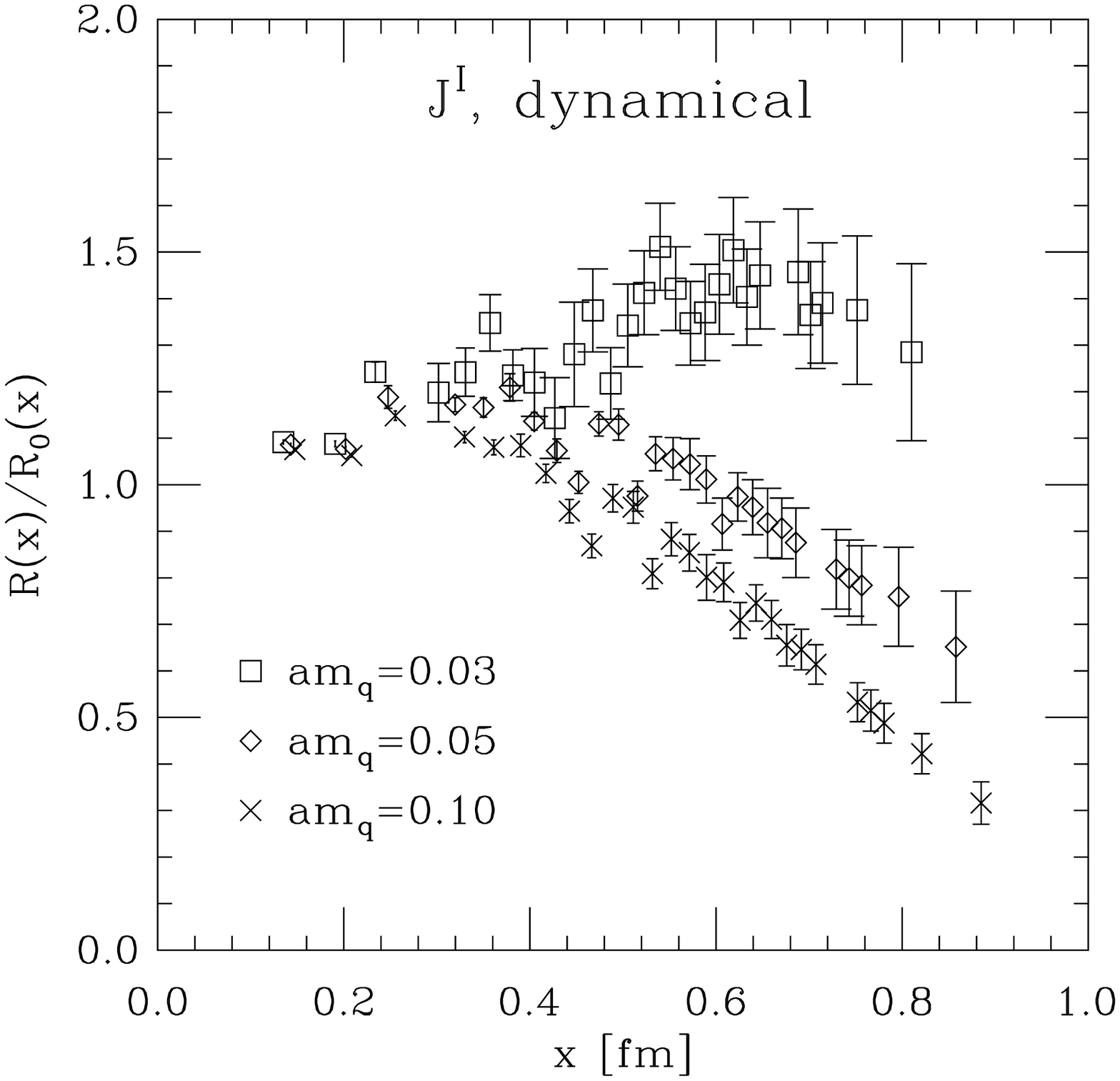}
\includegraphics[height=50mm,width=50mm]
{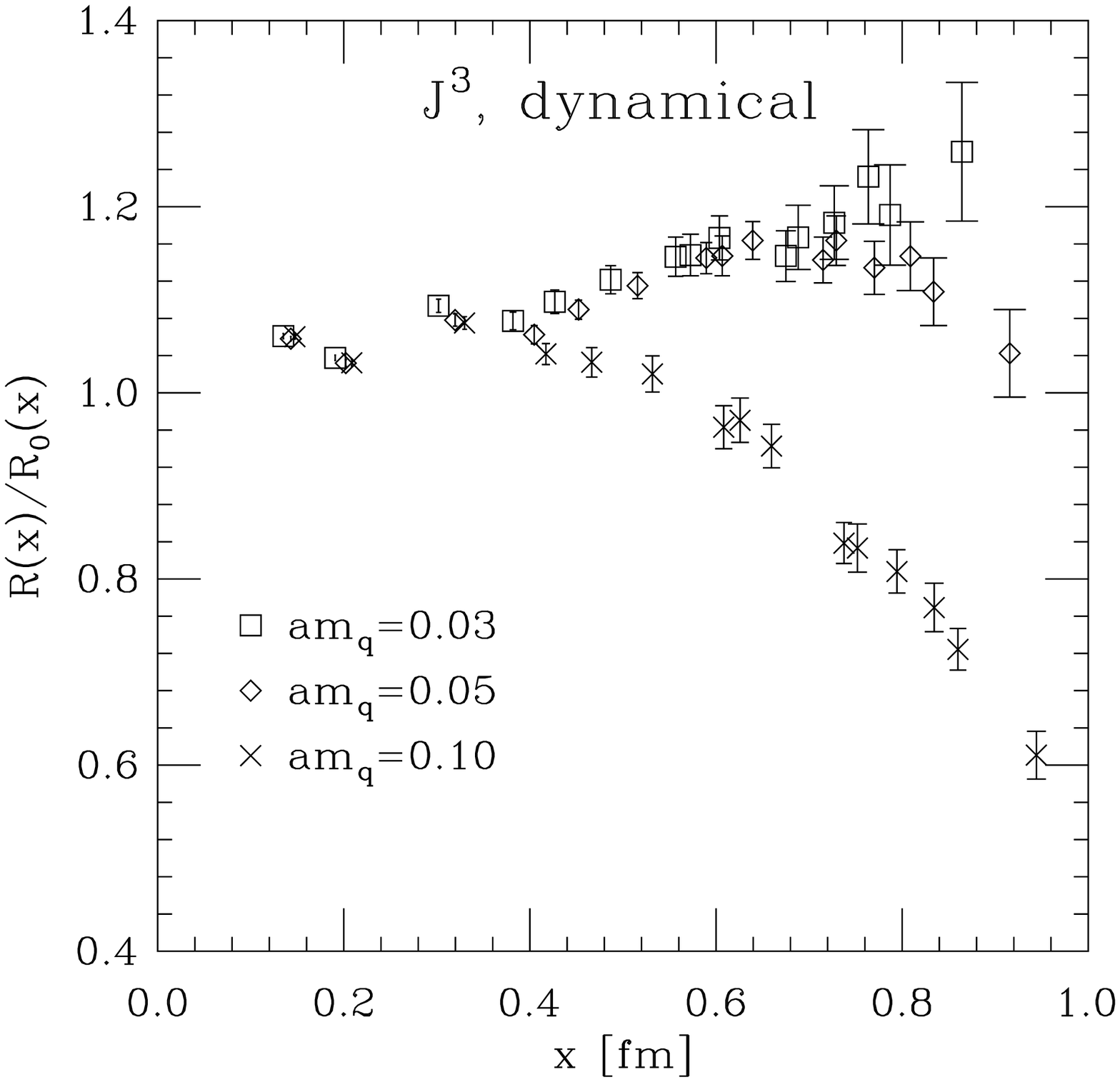}
\end{center}
\caption{Normalized correlation functions
$R(x)/R_0(x)$
for the four currents $J^5$, $J^{05}$, $J^I$ and $J^3$
from two flavor dynamical simulations for three quark
masses.}
\label{dyn}
\end{figure}

In Fig.~\ref{dyn}, we show the normalized correlation functions for all
four currents from our simulations with two flavors of dynamical fermions.

In the correlator for $J^5$, we see an attraction which
increases as the quark mass decreases. This attraction is weaker
and increases more slowly
than in the quenched data set, in Fig.~\ref{quen_all}. 

The $J^{05}$ correlator is repulsive at $am_q=0.10$. But as the 
quark mass decreases, 
it becomes less repulsive and at $am_q=0.03$, it is unity
in the range of $x$ that we consider. This mass dependence is consistent
with the quenched simulation result 
(see the second graph in Fig.~\ref{quen_all}). The dynamical result
confirms that the $J^{05}$ correlator behaves quite differently from the $J^{5}$
correlator although both currents contain a scalar diquark.

The $J^I$ correlator does not become negative at smaller quark masses; 
instead, it becomes more attractive. We saw similar behavior in the $Q=0$ sector of the quenched data set.
Here full QCD is unitary and the correlator remains positive.
The current $J^I$ vanishes in the nonrelativistic limit. Thus
it is not surprising
that the $J^I$ correlator is repulsive at big quark masses.
The attraction at small quark masses is a relativistic effect and we expect
it to become stronger as the quark mass becomes smaller.

The $J^3$ correlator, which involves the ``bad" diquark, 
is repulsive at
the largest quark mass, $am_q=0.10$. 
Then as the quark mass decreases, it
becomes slightly attractive. As in the quenched simulation, the attraction
in this correlator is much weaker than in the $J^5$ correlator.

\begin{figure}
\begin{center}
\includegraphics[height=50mm,width=50mm]
{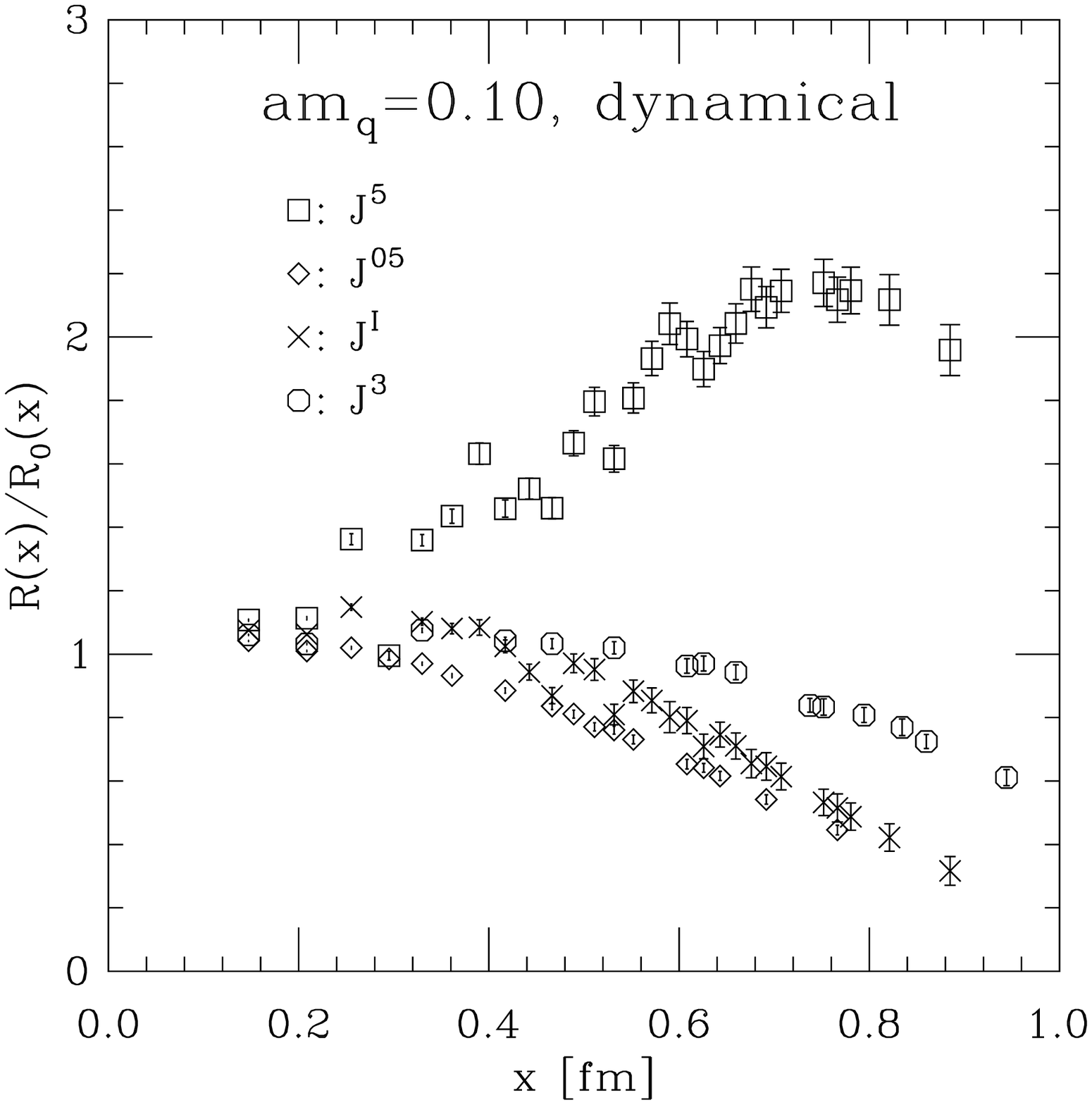}
\includegraphics[height=50mm,width=50mm]
{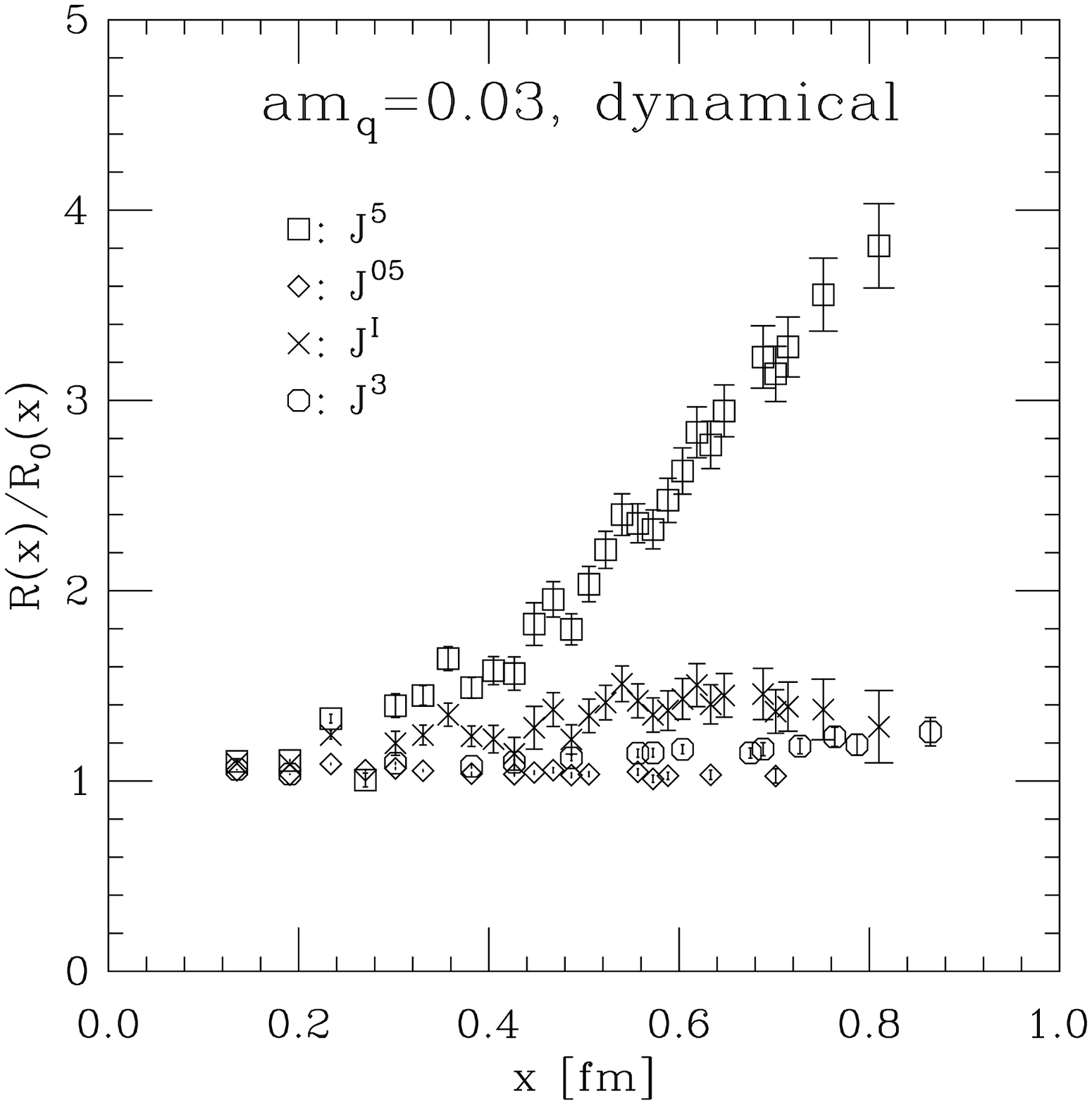}
\end{center}
\caption{Comparison of the four normalized correlators
$R(x)/R_0(x)$ from the dynamical simulation.
The left graph is for the largest quark mass. The right
for the lowest quark mass.}
\label{dyn_4J}
\end{figure}

In Fig.~\ref{dyn_4J}, we put all four correlators together for
comparison. The left graph is for the largest quark mass $am_q=0.10$.
The right graph is for the lowest quark mass $am_q=0.03$.
When the quark mass is big, 
only the $J^5$
correlator is attractive. The other three are repulsive
as we can see in the left graph in Fig.~\ref{dyn_4J}.
As the quark mass becomes small in
the dynamical simulation, 
the other three correlators turn into attractive 
as is shown in the right graph in Fig.~\ref{dyn_4J}.
However the attraction in
the $J^5$ correlator is still much stronger than in the other three.
We also observe a slightly stronger attraction in the
$J^I$ correlator than in the $J^{05}$ and $J^3$ correlators.

\begin{figure} 
\begin{center}
\includegraphics[width=50mm,height=50mm]{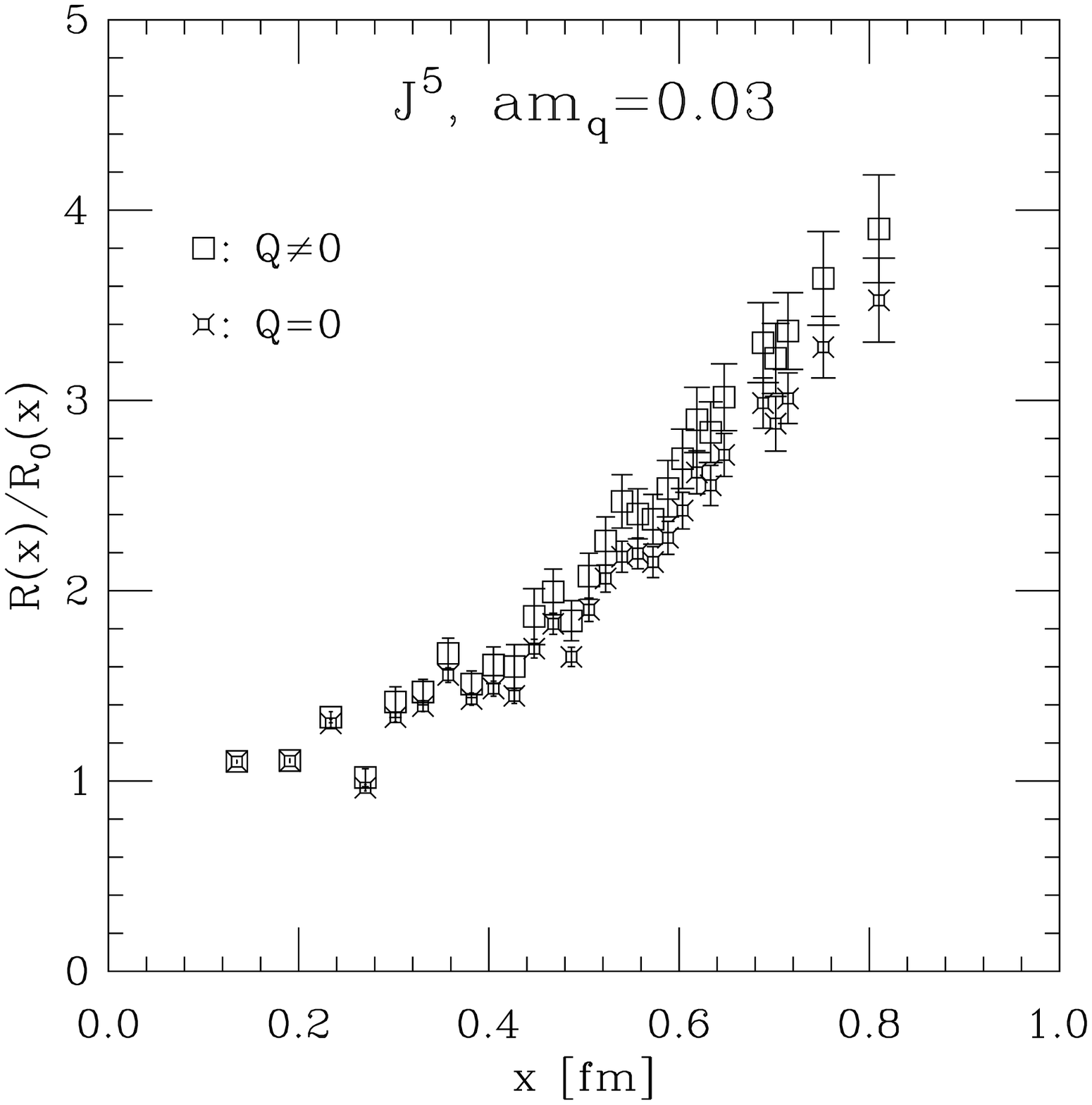}
\includegraphics[width=50mm,height=50mm]{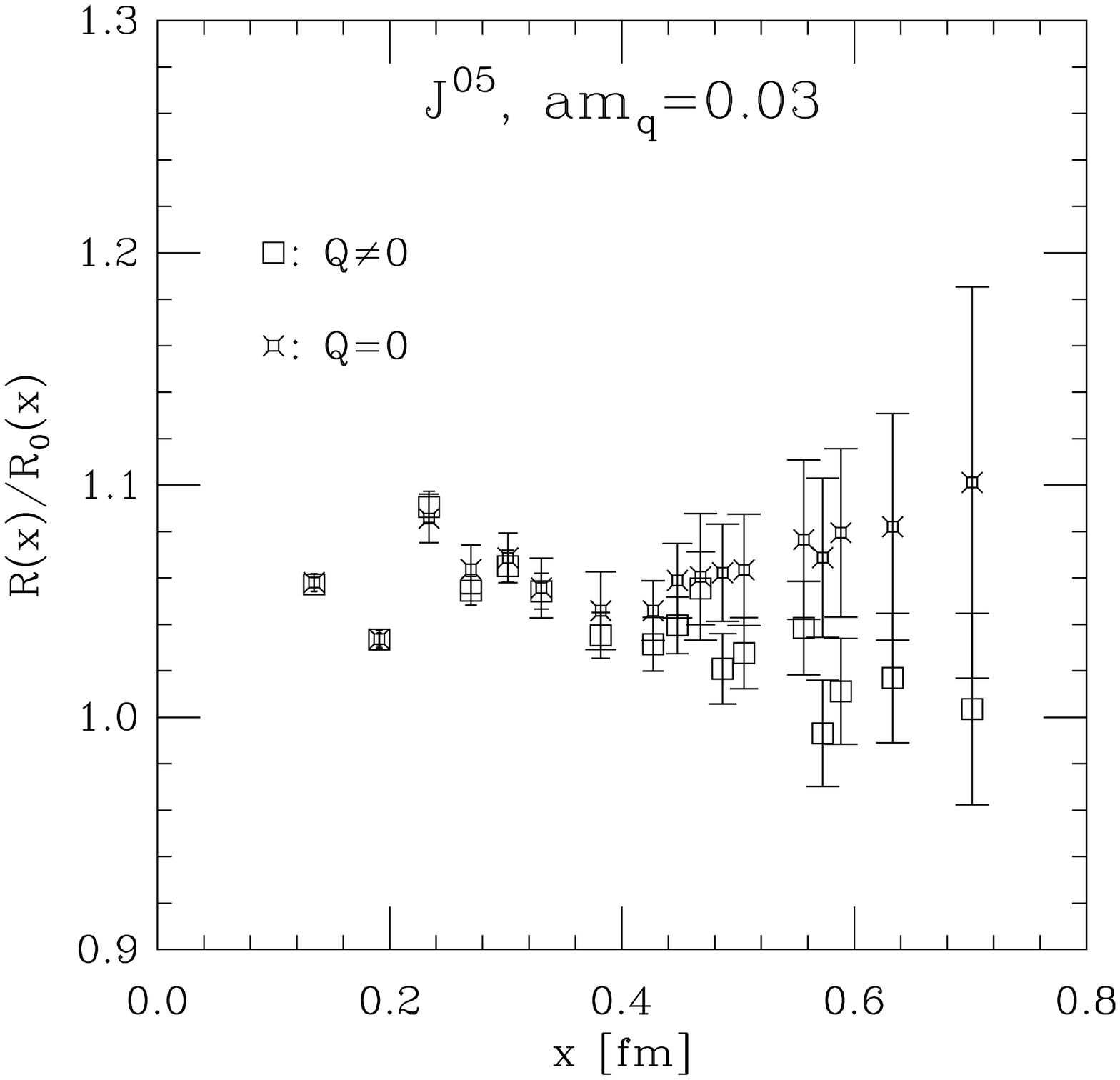}\\
\includegraphics[width=50mm,height=50mm]{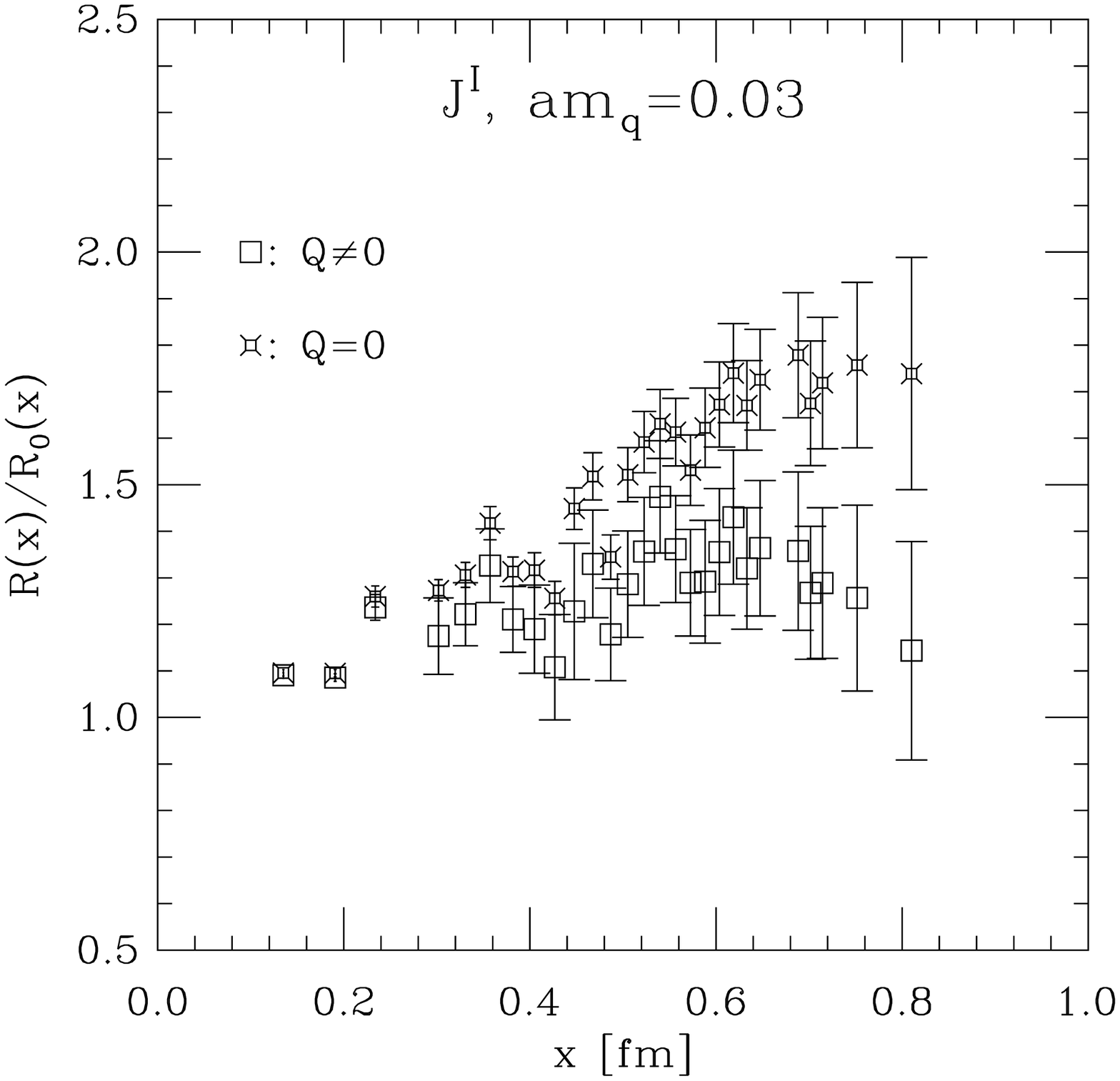}
\includegraphics[width=50mm,height=50mm]{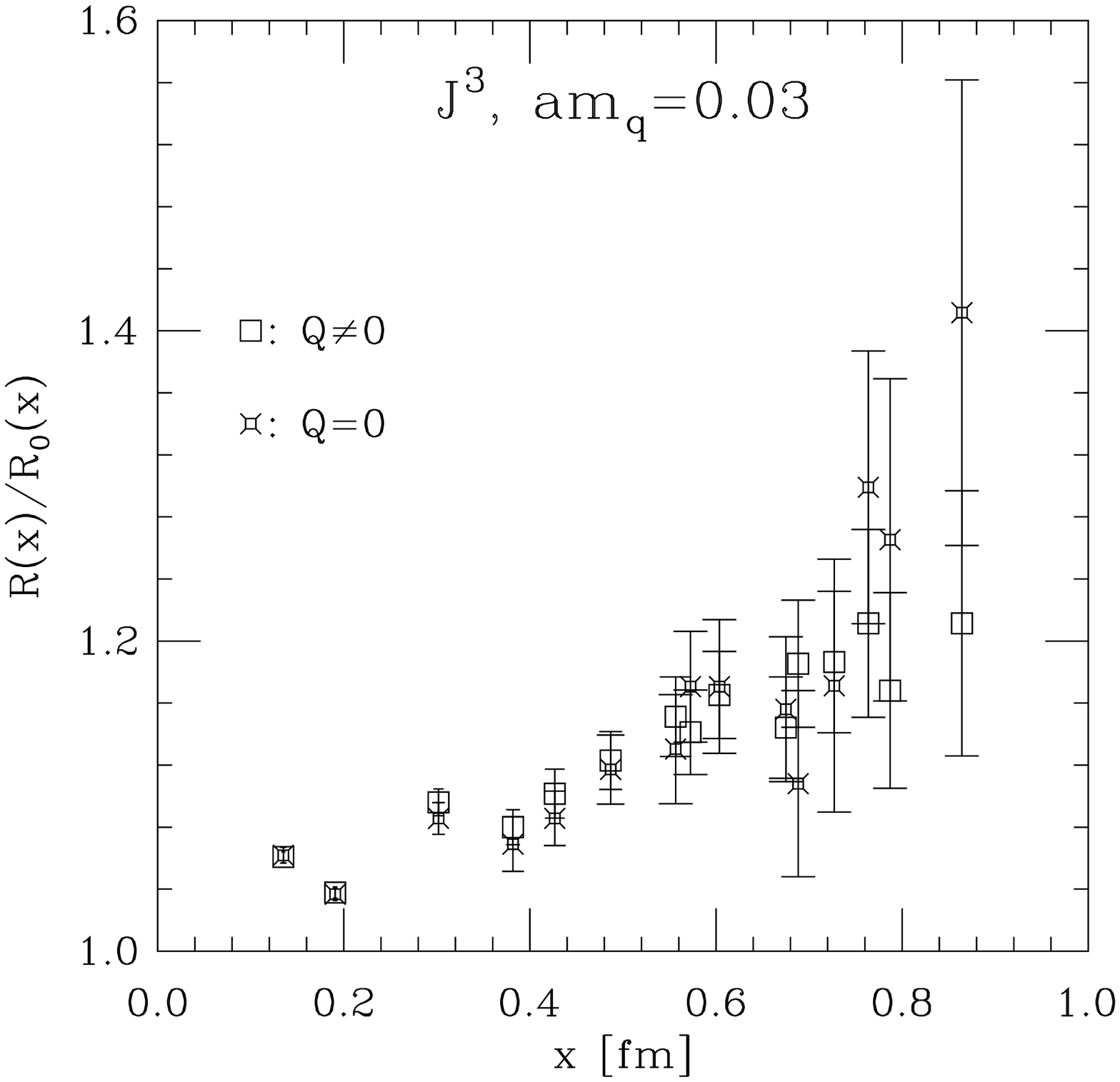}
\end{center}
\caption{Normalized correlators for currents $J^5$, $J^{05}$, $J^I$ and $J^3$ from $N_f=2$ simulations
at quark mass $am_q=0.03$.
The correlators
from configurations
with $Q\neq0$ are compared with
those from configurations with $Q=0$.}
\label{dyn0.03Q}
\end{figure}
\begin{figure}
\begin{center}
\includegraphics[width=50mm,height=50mm]{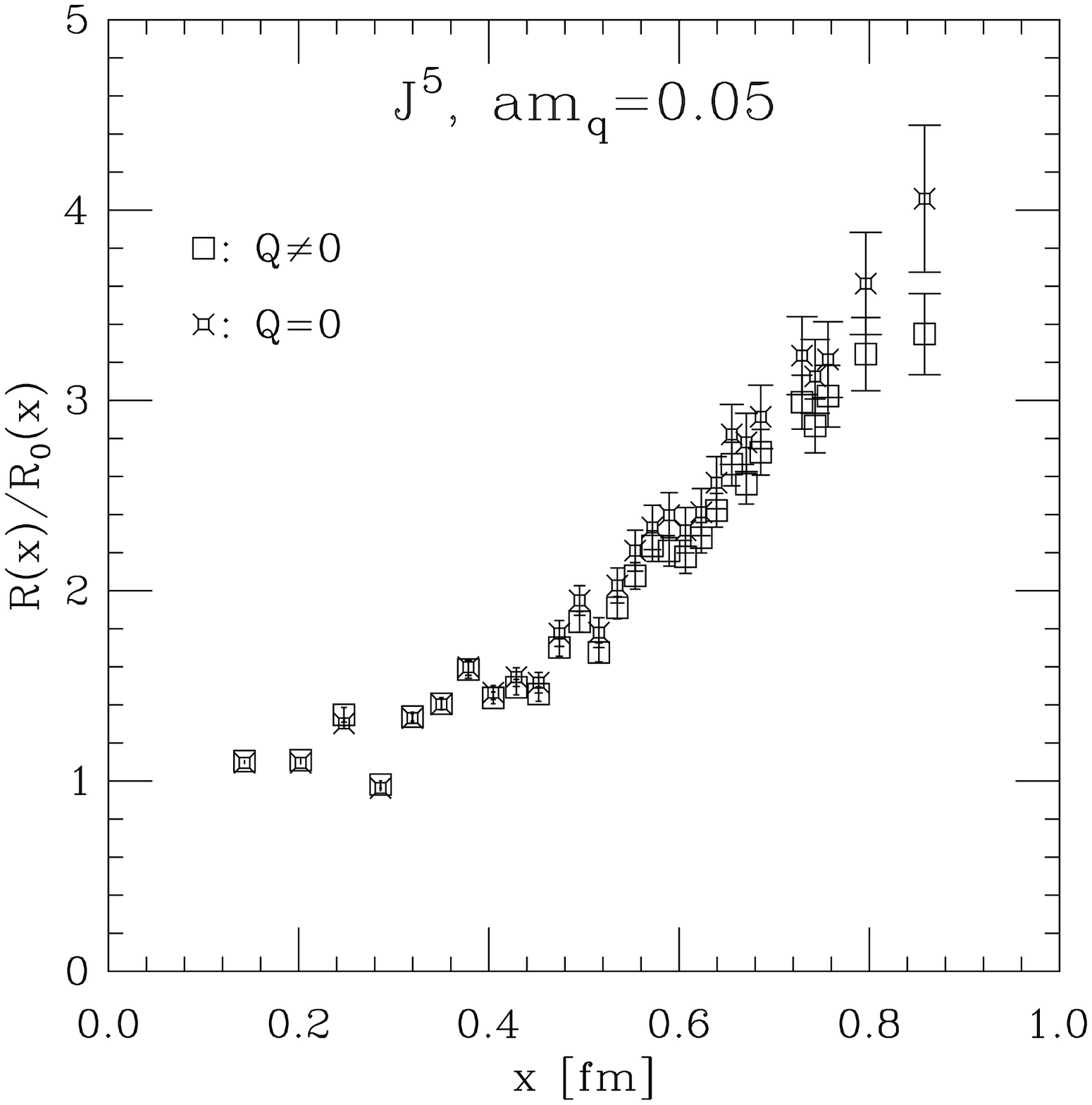}
\includegraphics[width=50mm,height=50mm]{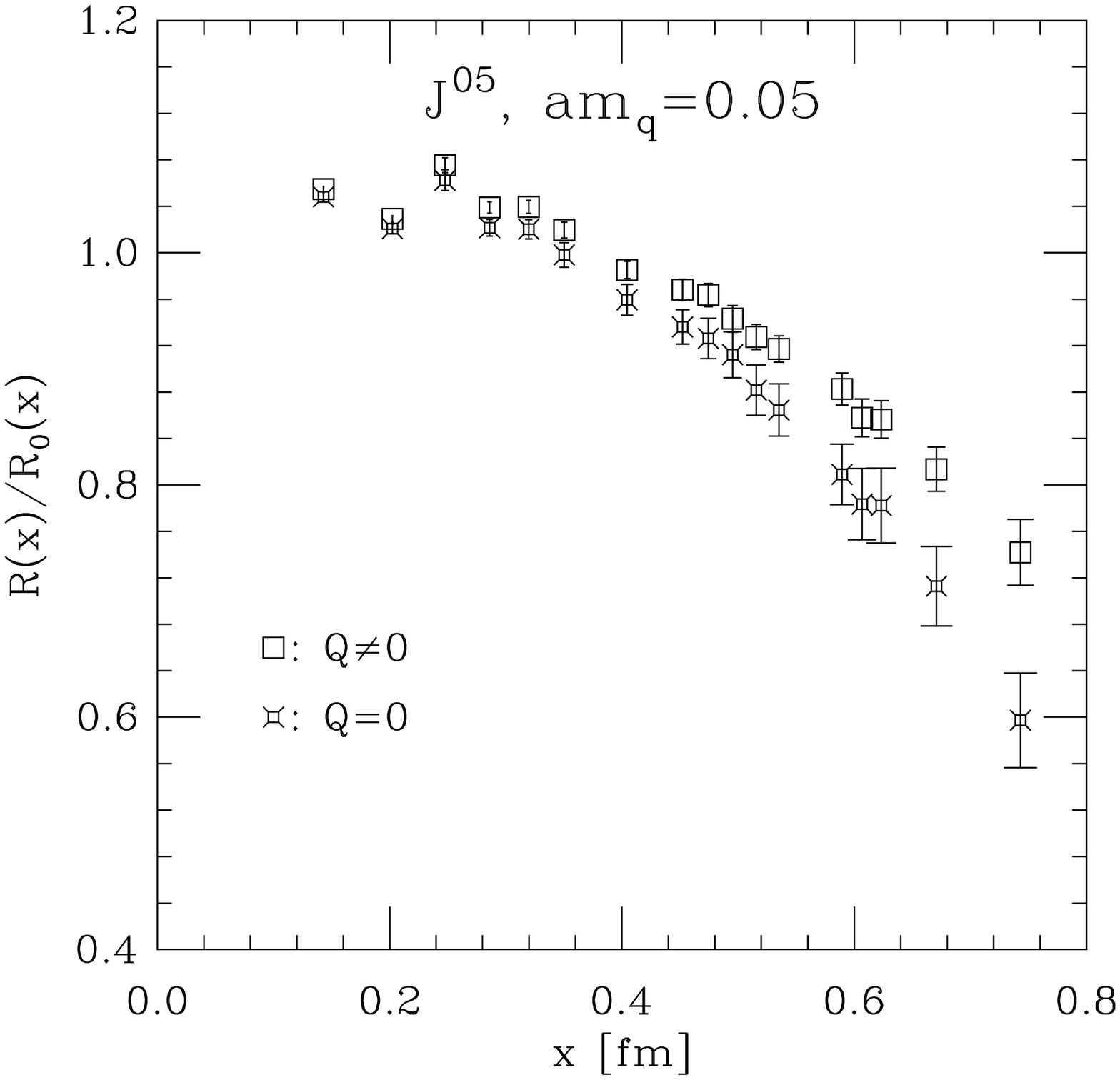}\\
\includegraphics[width=50mm,height=50mm]{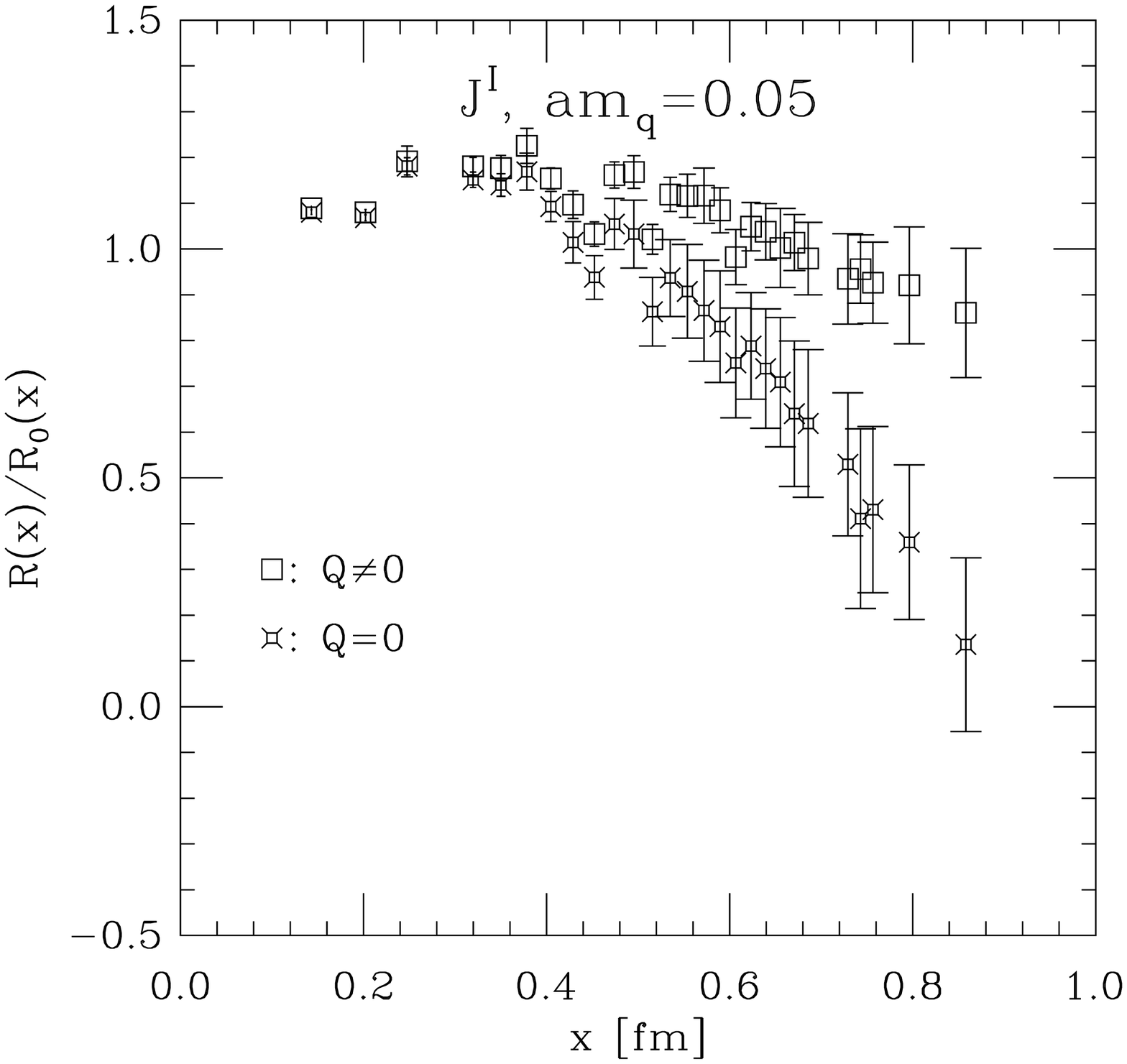}
\includegraphics[width=50mm,height=50mm]{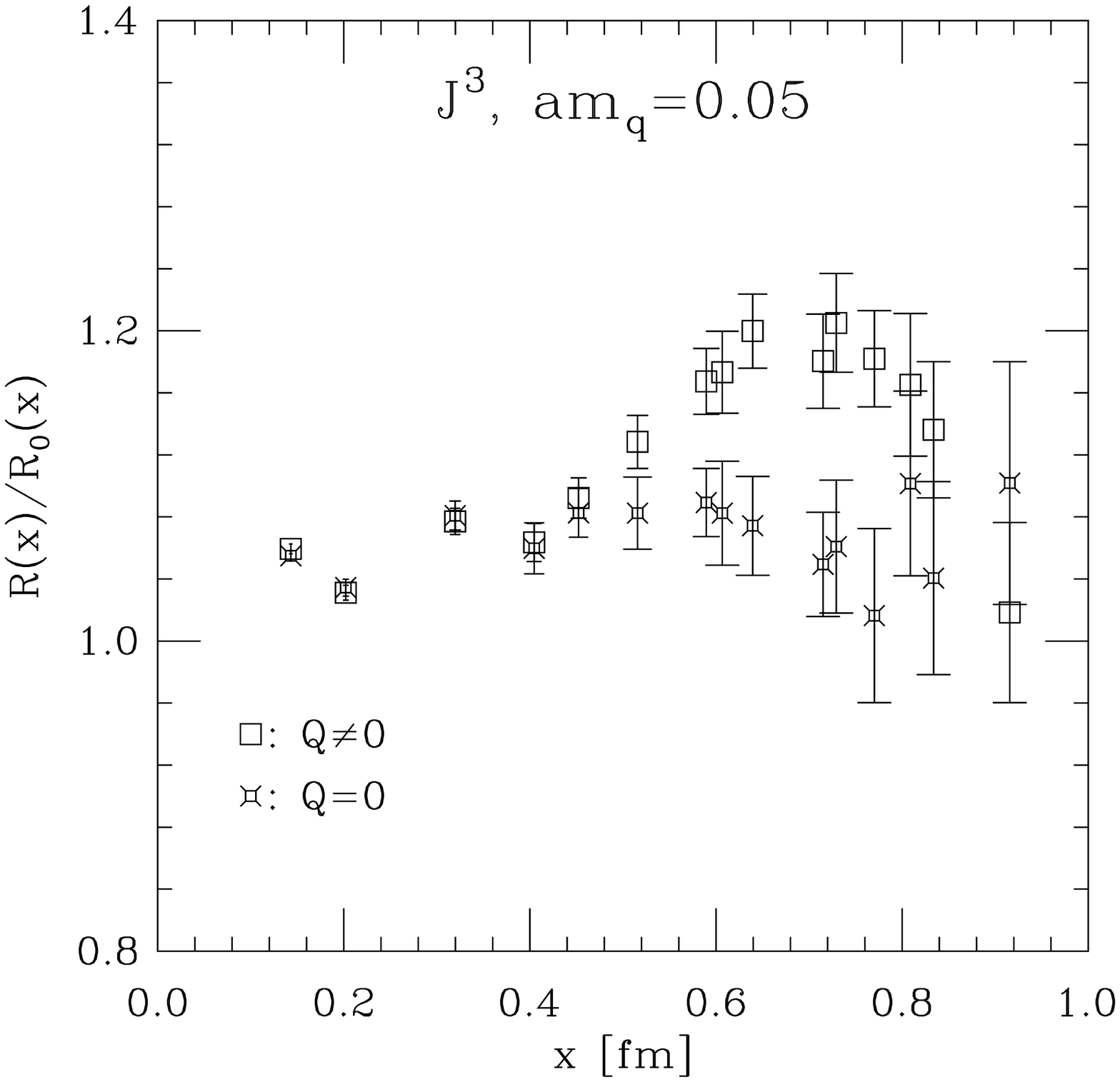}
\end{center}
\caption{Normalized correlators for currents $J^5$, $J^{05}$, $J^I$ and $J^3$
from $N_f=2$ simulations
at quark mass $am_q=0.05$.
The correlators
from configurations
with $Q\neq0$ are compared with
those from configurations with $Q=0$.}
\label{dyn0.05Q}
\end{figure}
\begin{figure}
\begin{center}
\includegraphics[width=50mm,height=50mm]{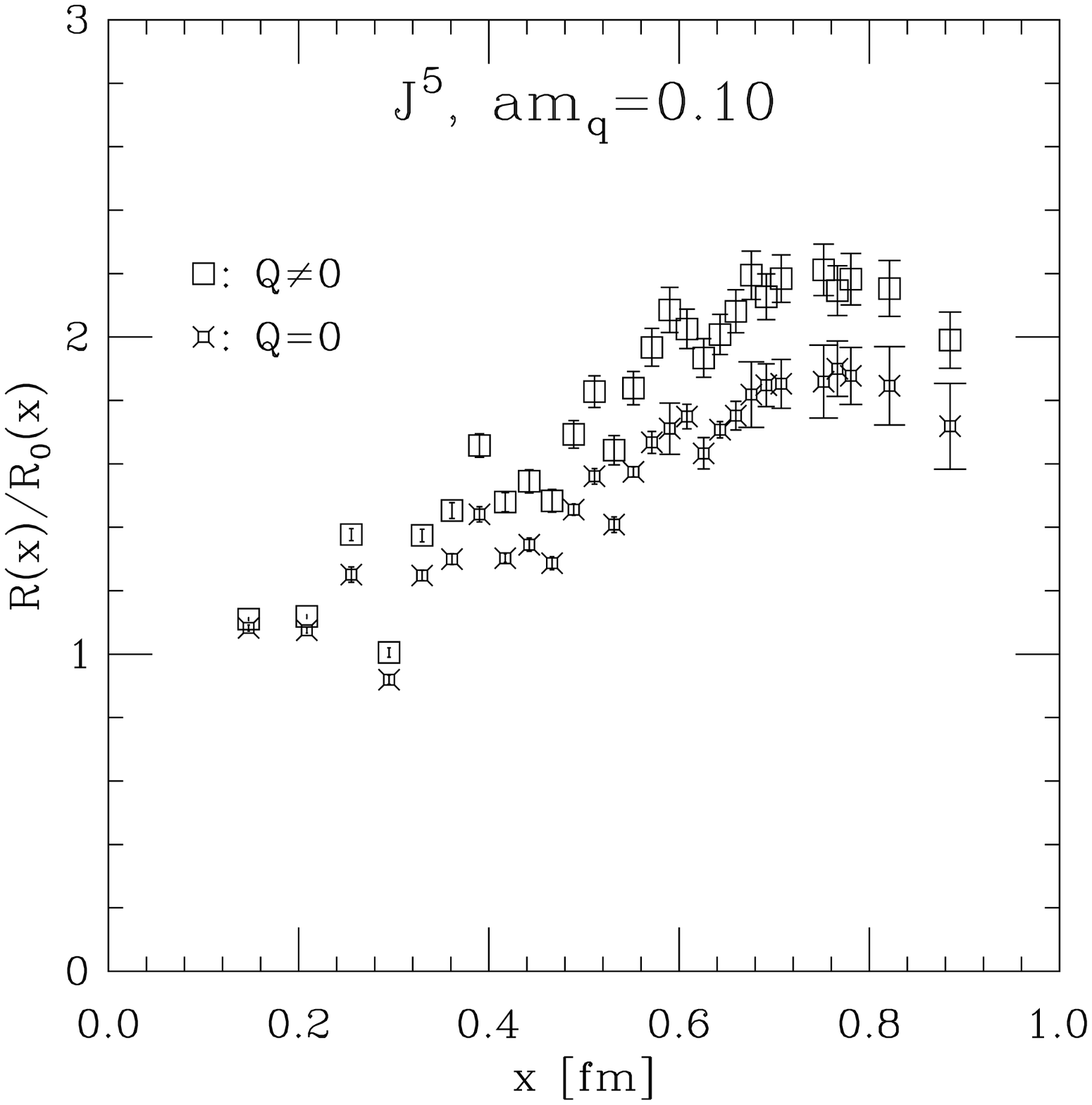}
\includegraphics[width=50mm,height=50mm]{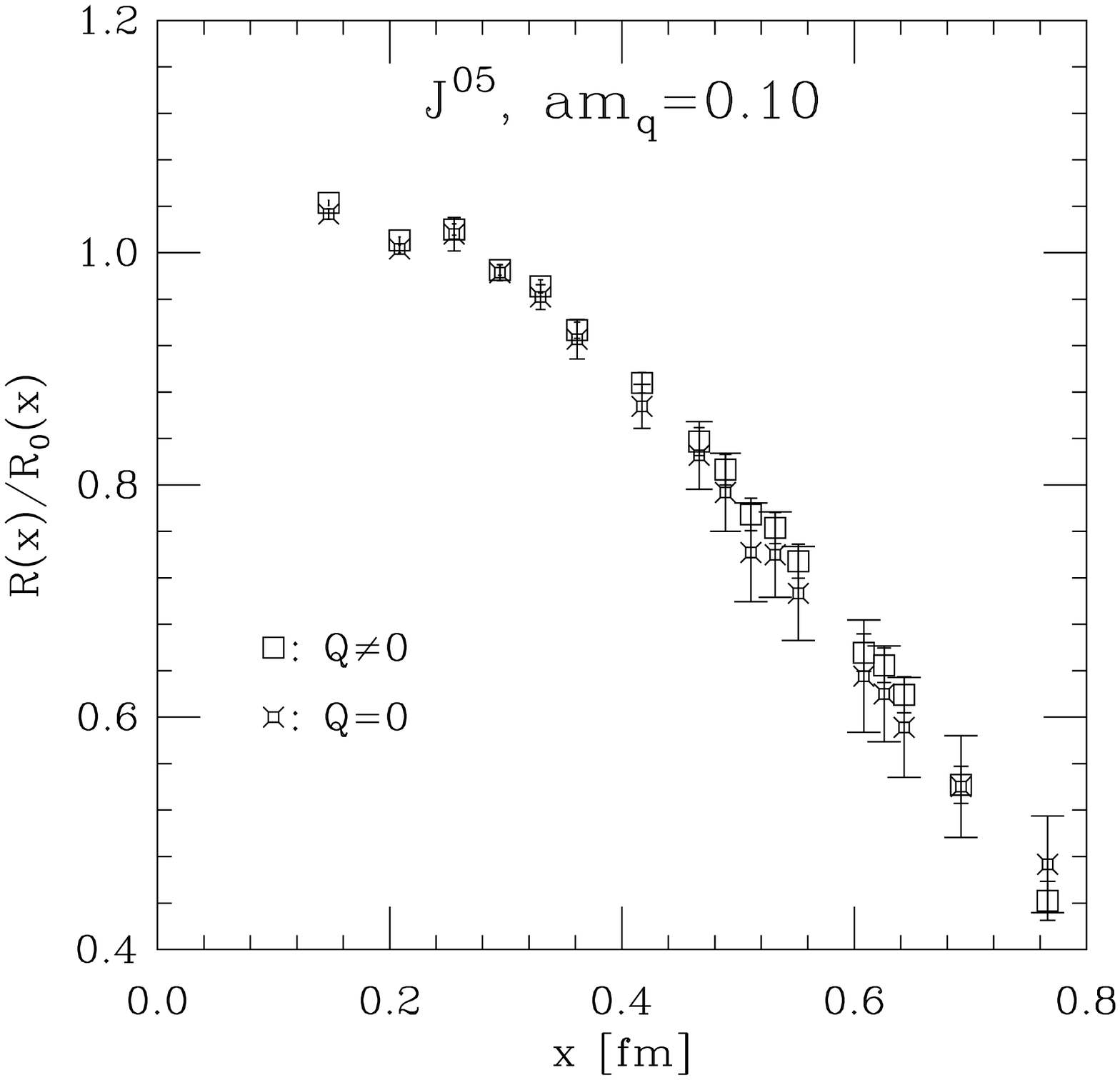}\\
\includegraphics[width=50mm,height=50mm]{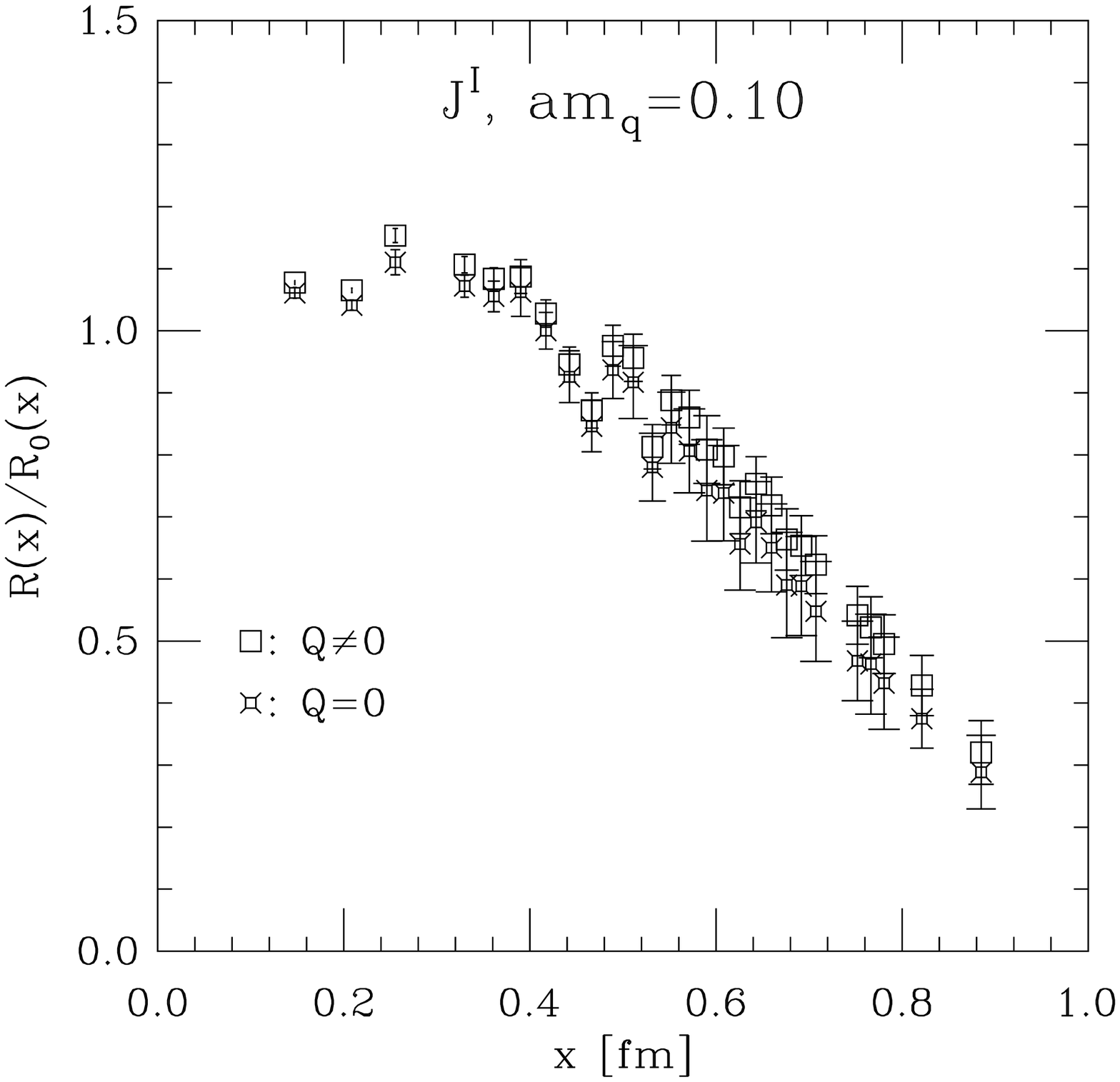}
\includegraphics[width=50mm,height=50mm]{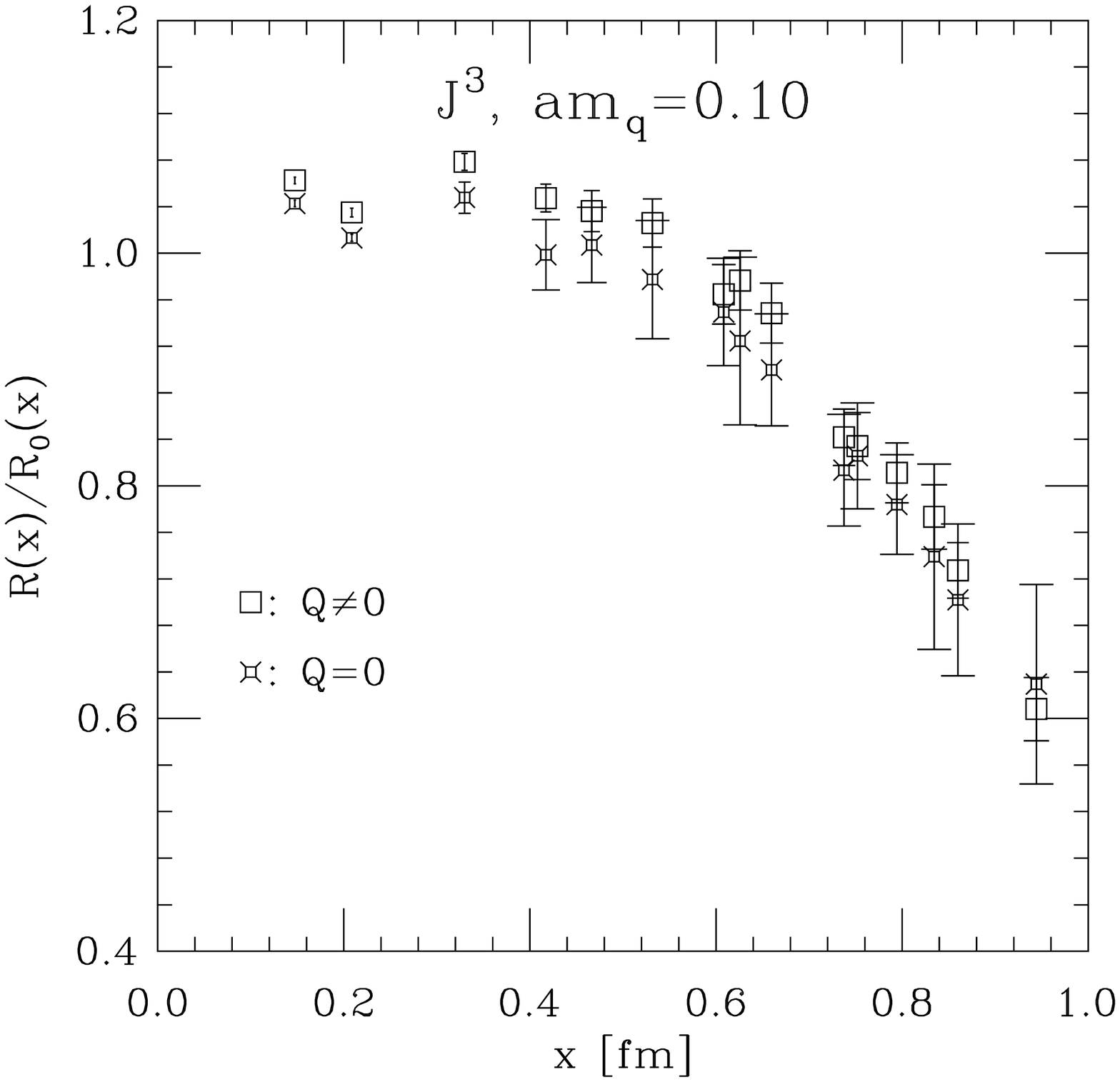}
\end{center}
\caption{Normalized correlators for currents $J^5$, $J^{05}$, $J^I$ and $J^3$
from $N_f=2$ simulations
at quark mass $am_q=0.10$.
The correlators
from configurations
with $Q\neq0$ are compared with
those from configurations with $Q=0$.}
\label{dyn0.10Q}
\end{figure}
As we did in the analysis of the quenched data, we can separate our 
dynamical configurations according to their topological charges.
We have 60 configurations for $am_q=0.10$ and 80 configurations
for both $am_q=0.05$ and 0.03. The numbers of high $Q$ ($|Q|\ge2$) configurations
for the three quark masses are 37, 33 and 16 respectively. Clearly,
high $Q$ configurations are suppressed as the quark mass falls.

In Fig.~\ref{dyn0.03Q}, \ref{dyn0.05Q} and \ref{dyn0.10Q}, 
the four correlators from 
$Q\neq0$ configurations
are compared with their counterparts 
from zero $Q$ configurations 
for the three quark masses. In contrast to the quenched data set, the effects of topology
are small. (Note that not even the relative sizes of the correlators
from  different values of $|Q|$ are the same for all three masses.)
 The only place where topology appears to play a big role is in the $J^I$ correlator 
at the smallest quark mass. Even there, the large-$Q$ correlator 
remains positive over the entire measured $x$ range. Based on our quenched results, 
we might have imagined a scenario in which the $Q\ne0$ $J^I$ correlator was negative, but the correlator
summed over all $Q$ was positive, as required by unitarity. That does not seem to be what is going on.
Of course, the physics which is responsible for the probability of encountering a particular value of
$Q$ in an ensemble is almost certainly different in quenched and full QCD, so maybe it is not surprising that the
correlation functions are different.

\section{Summary and conclusion}\label{Summ}
We studied the point to point light baryon correlators for four currents
using quenched and two flavor dynamical overlap simulations.
As far as we know, this is the first test of point to point correlators
for light baryons in full QCD.
In these four currents, there are three diquark structures: scalar ($J^5$
and $J^{05}$),
pseudoscalar ($J^I$) and axial vector ($J^3$).

These correlators are rather different in quenched
approximation and in full QCD. The quenched $J^5$ correlator is more attractive than in full QCD,
and the quenched $J^I$ correlator is negative at small quark mass.
These features are correlated with the presence of topological zero modes in the quenched data set.

We have not attempted to make a direct comparison with models. We are aware that in the instanton liquid model,
baryon channels with different diquark quantum numbers
show a strong response to the presence of topological zero modes
\cite{Schafer:1995uz,Schafer:1994nv,Cristoforetti:2004kj}.
It might be, that quenched results could be a diagnostic for models even though
they are not reliable indicators of the behavior of correlators in full QCD.

The quark masses we are simulating are interesting ones
because they go from a region where qualitatively speaking
chiral symmetry effects are not so important ($\sim$100 MeV),
to a region where they are probably important (30-40 MeV).
As our quark mass falls, the dynamical results show apparent
changes in each correlator (see Fig.~\ref{dyn_4J}).
In this quark mass range,
the scalar diquark created by $\gamma_5$ seems to play a greater role in baryon correlators, than the others.
The $J^5$ correlator is the most attractive one, and the attraction in this channel rises 
strongly
as the quark mass falls.

An attraction appears in the correlator of
$J^I$ 
at the lowest quark mass in the dynamical simulation. In the 
light-light-heavy system~\cite{Alexandrou:2006cq}, 
the pseudoscalar diquark was shown to be
heavier than the axial vector and scalar diquarks. Here
the attraction in the $J^I$ correlator is slightly stronger 
than in the $J^{05}$
and $J^3$ correlators. Diquark
effects in light baryons are apparently different from those in heavy baryons.

The big difference between the $J^{05}$ and $J^5$ correlators is also a strong
statement that diquark correlations are not a dominant factor in light
baryons. Both correlators contain scalar diquarks. 
The difference between these correlators is how the diquark couples to the third quark,
and this coupling apparently matters: we
 are not seeing an isolated diquark.

Finally, we have to conclude by again listing the shortcomings in this project:
even our lightest quark masses are quite a bit heavier than the physical up and down quark masses.
Simulations at smaller quark mass will require larger volume to prevent an effective restoration of chiral symmetry
as the quark mass is reduced.
Given the lattice artifacts we have seen in the quenched part of our study,
 we recommend against using mixed action or partially
quenched simulations for these studies.
\section*{Acknowledgments}
This work was supported by the US Department of Energy,
French ANR under
grant ANR-05-CIGC-001-09 and by the Deutsche
Forschungsgemeinschaft in the SFB/TR 09. 
We thank the DESY computer 
center in Zeuthen for essential computer
resources and support.
This work was in part based on the MILC collaboration's 
public lattice gauge theory code \cite{MILC}.

\end{document}